\documentclass[11pt,draftclsnofoot,peerreview]{IEEEtran}
\usepackage{graphicx,times,amsmath}

\newtheorem{corollary}{Corollary}[section]
\newtheorem{lemma}{Lemma}[section]
\newtheorem{theorem}{Theorem}[section]
\newtheorem{claim}{Claim}

\title{Flow-optimized Cooperative Transmission for the Relay
  Channel\thanks{This research is supported in part by the National
    Science Foundation under Grant CNS-0626863.}  }

\author{\emph{Tan F. Wong$^\dag$, Tat M. Lok$^\ddag$, and John M. Shea$^\dag$} \\~\\
  $\dag$Department of Electrical and Computer Engineering \\
  University of Florida \\
  Gainesville, FL 32611-6130, U.S.A. \\
  Tel: +1-352-392-2665~~~
  Fax: +1-352-392-0044 \\
  Email: \texttt{\{twong,jshea\}@ece.ufl.edu} \\~\\
  $\ddag$Department of Information Engineering \\
  The Chinese University of Hong Kong \\
  Shatin, Hong Kong \\
  Tel: +852-2609-8455~~~
  Fax: +852-2603-5032 \\
  Email: \texttt{tmlok@ie.cuhk.edu.hk}}

\begin{document}
\maketitle

\begin{abstract} 
  This paper describes an approach for half-duplex cooperative
  transmission in a classical three-node relay channel. Assuming
  availability of channel state information at nodes, the approach
  makes use of this information to optimize distinct flows through the
  direct link from the source to the destination and the path via the
  relay, respectively. It is shown that such a design can effectively
  harness diversity advantage of the relay channel in both high-rate
  and low-rate scenarios. When the rate requirement is low, the
  proposed design gives a second-order outage diversity performance
  approaching that of full-duplex relaying. When the rate requirement
  becomes asymptotically large, the design still gives a
  close-to-second-order outage diversity performance.  The design also
  achieves the best diversity-multiplexing tradeoff possible for the
  relay channel. With optimal long-term power control over the fading
  relay channel, the proposed design achieves a delay-limited rate
  performance that is only $3.0$dB ($5.4$dB) worse than the capacity
  performance of the additive white Gaussian channel in low- (high-)
  rate scenarios.
\end{abstract}

\pagebreak

\section{Introduction}
It is well known that the performance of a wireless network can be
significantly improved by cooperative transmission among nodes in the
network.  Many cooperative transmission designs aim to exploit
cooperative diversity that is inherently present in the network.  Such
designs have been suggested in \cite{Sendonaris03a,Sendonaris03b} for
cellular networks.  Recently there has been much interest in achieving
cooperative diversity in a classical three-node relay channel
\cite{Cover79}, which represents the simplest wireless network that
can derive advantages from cooperative transmission.

The relay channel has been thoroughly studied in \cite{Cover79}.
Bounds on the capacity have been given for the general relay channel,
and the capacity has been calculated for the special case of degraded
relay channels. The coding techniques suggested in \cite{Cover79}
assume that the relay can operate in a full-duplex manner; i.e., it
can transmit and receive at the same time. It is commonly argued that
full-duplex operation is not practical for most existing wireless
transceivers. Thus the restriction of half-duplex operation at the
relay is usually considered in cooperative transmission designs.

Since the relay cannot transmit and receive simultaneously, a
time-division approach is employed in half-duplex relaying
\cite{Laneman04}. The source first transmits to the destination, and
the relay listens and ``captures'' \cite{Host05} the transmission from
the source at the same time. Then the relay aids the transmission by
sending processed source information to the destination. Note that the
source may still send data to the destination when the relay
transmits. Several techniques to process and forward the received data
by the relay have been suggested. These techniques include the
decode-and-forward (DF) and amplify-and-forward (AF) approaches
\cite{Laneman04}. In the DF approach, the relay decodes the received
signal from the source and then forwards a re-encoded signal to the
destination. In the AF approach, the relay simply amplifies and
forwards the signal received from the source to the destination.

The performance of the DF approach is limited by the capability of the
relay to correctly decode the signal received from the source. This in
turn depends on the quality of the link from the source to the relay.
On the other hand, the AF approach performs poorly in low
signal-to-noise ratio (SNR) situations in which the relay forwards
mainly noise to the destination. In addition, the time-division
approach leads to rate losses that are significant when the relay
channel is to support high rates. Some enhanced versions of the AF and
DF approaches have been proposed to solve the rate loss problem. A
distributed space-time-coding protocol is developed in
\cite{Laneman03}. An incremental AF technique which requires feedback
from the destination to the source is developed in \cite{Laneman04}.
The non-orthogonal AF and dynamic DF techniques suggested in
\cite{Azarian05} allow the duration of the relay listening to the
transmission from the source to adapt to the condition of the link
from the source to the relay. In particular, the dynamic DF technique
is shown to be superior to all the cooperative diversity techniques
(except perhaps the incremental relaying techniques) mentioned above.
A bursty AF technique is also suggested in \cite{Avestimehr06} to
solve the noise forwarding problem of the AF approach when the SNR is
low. It is shown that the bursty AF technique achieves the best outage
performance at the asymptotically low SNR regime. We note that all
these cooperative diversity techniques mentioned so far are designed
with the constraint that channel state information is not available at
the source and the relay. Some practical code designs for the DF and
space-time-coding approaches have been suggested in \cite{Valenti03}
and \cite{Janani04}, respectively.

When the links in the relay channel suffer from slow fading, it is
conceivable that the channel state information (or at least the
channel quality information) can be estimated and passed to the nodes.
The source and relay may then use this information to optimize the
cooperative protocol to achieve better performance. Such a design has
been considered in \cite{Host05}, in which optimal power control is
performed at the source and relay in order to maximize the ergodic
rates achieved by the DF and compress-forward approaches.

In this paper, we assume that the channel state information is
available, and we develop time-division cooperative diversity designs
that perform well in both high-rate and low-rate scenarios. The main
distinguishing feature of the proposed approach, compared with the
cooperative designs mentioned above, is that we do not employ the
approach of the relay ``capturing'' the transmission from the source
to the destination. Instead, we divide the information to be sent to
the destination into three flows. The source employs cooperative
broadcasting \cite{Cover72,Bergmans74} to intentionally send two
\emph{distinct} flows of data to the relay and destination,
respectively, in the first time slot. The relay helps to forward, in
the DF manner, the data that it receives to the destination in the
second time slot, during which the source concurrently sends the
remaining flow of data to the destination. The transmit powers of the
source and relay as well as the durations of the time slots are
optimized according to the link conditions and the rate requirement.
This constitutes a form of optimal flow control.

Due to the DF nature of the proposed design, there is an implicit
restriction on the decoding delay.  Thus we will employ the
capacity-versus-outage framework \cite{Ozarow94,Biglieri98} to
evaluate the performance of the proposed design.  We will show that
the proposed design can efficiently achieve cooperative diversity in
both high-rate and low-rate scenarios. In particular, when the rate
requirement is asymptotically small, the outage performance of the
proposed design approaches that of full-duplex relaying with DF,
giving a second-order diversity performance. On the other hand, when
the rate requirement is asymptotically large, the proposed approach
still gives a close-to-second-order diversity performance. Moreover,
the design also gives the best diversity-multiplexing tradeoff
\cite{Zheng03} possible for the relay channel. Together with the
application of optimal long-term power control \cite{Caire99}, the
design can give very good delay-limited rate performance again in both
low-rate and high-rate scenarios.

We note that the two basic building blocks for the proposed approach
are cooperative broadcasting (CB) in the first time slot and multiple
access (MA) in the second time slot. The combination of CB and MA
allows distinct flows of data be sent through the relay and
through the direct link from the source to the destination,
and hence can be viewed as a generalized form of routing.
A practical advantage of the proposed design is that the basic
building blocks are the well known CB and MA approaches. Practical MA
coding designs have been well studied, e.g.  see
\cite{Verdu98,Wang99}, while practical CB coding designs are currently
available \cite{Yu05}--\nocite{Airy04}\cite{Erez05}.

\section{Relay Channel: Full-Duplex Bounds}
\label{se:fd}
\begin{figure}
  \centering
  \includegraphics[width=0.7\textwidth]{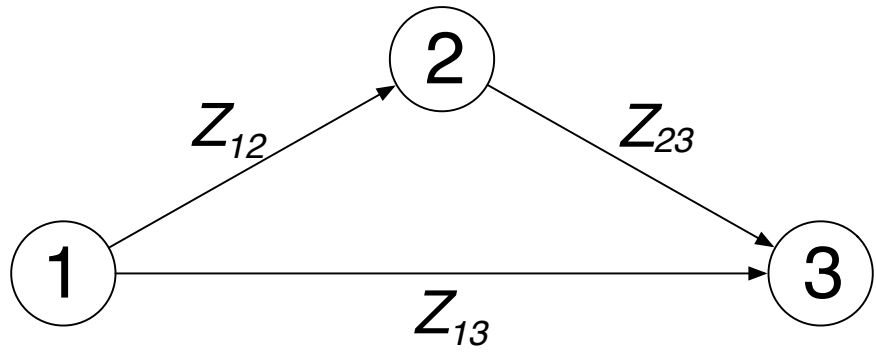}
  \caption{The classical 3-node relay channel.}
  \label{f:relay}
\end{figure}

Consider a classical three-node relay network, which consists of a
source node 1, a relay node 2, and a destination node 3 as shown in
Fig.~\ref{f:relay}.  We assume that each link in the figure is a
bandpass Gaussian channel with bandwidth $W$ and one-sided noise
spectral density $N_{0}$. Let $Z_{ij}$ denote the power gain of the
link from node $i$ to node $j$.  The link power gains are assumed to
be independent and identically distributed (i.i.d.) exponential random
variables with unit mean. This corresponds to the case of independent
Rayleigh fading channels with unit average power gains. The results in
the sequel can be easily generalized to include the case of
non-uniform average power gains.

In this section, we consider the case in which the relay node is
capable of supporting full-duplex operation. Our goal is to support an
information rate\footnote{Strictly speaking, the word ``rate'' here
  should be replaced by ``spectral efficiency'', since the unit
  involved is nats/s/Hz. Nevertheless we will use the terminology
  ``rate'' throughout this paper for convenience.} of $K$ nats/s/Hz
from the source to the destination.  We assume that the link power
gains change slowly so that they can be estimated, and hence the power
gain information is available at all nodes. The source and relay nodes
can make use of this information to control their respective transmit
power and time so that the total transmit energy is minimized.  For
convenience, we consider a slotted communication system with
unit-duration time slots. Let $P_t$ be the total transmit energy of
the source and relay needed to support the transmission of $KW$ nats
of information from the source to the destination in a time slot.
Since the duration of a time slot is one, $P_t$ is also the total
average transmit power required. Although interpreting $P_t$ as the
total average power may not carry any significant physical meaning, it
is customary to speak of ``power'' rather than ``energy'' in for
communication engineers.  Unless otherwise stated, we will hereafter
consider a normalized version of $P_t$, namely the rate-normalized
overall signal-to-noise ratio (RNSNR) of the network:
\[
S \stackrel{\triangle}{=} \frac{P_t}{N_0 W} \frac{1}{e^K -1}.
\] 
The RNSNR can be interpreted as the additional SNR, in dB, needed to
support the required rate of $K$ nats/s/Hz, in excess of the SNR
required to support the same rate in a simple Gaussian channel with
unit gain.  This normalization is convenient as we will consider
asymptotic cases when $K$ approaches zero and infinity.

We note that the use of the RNSNR to characterize our results has two
important implications. First, since the total transmit energy of the
source and relay is used in defining the RNSNR, no individual power
limits are put on the source and relay.  The results in this paper can
be viewed as bounds if additional individual power limits are imposed.
Our choice of focusing on the total energy comes from the viewpoint
that the relay channel considered forms a small component of a larger
wireless network. In this sense, it is fairer to compare the total
transmit energy incurred in sending information from the source to the
destination by employing cooperative diversity to that incurred in
direct transmission.  Second, the normalization by the factor $e^K -1$
implies that the additional SNR in dB to combat fading can only be a
\emph{constant} over the SNR required to achieve the target rate in a
Gaussian channel, regardless of the rate requirement. That is, we
restrict the SNR to increase at the same rate as in a Gaussian channel
to cope with increases in the transmission rate through the relay
channel. In a sense, this restriction enforces efficiency of energy
usage.

Employing well known capacity bounds on the relay channel
\cite{Host05,Cover79,Cover91}, we can obtain the following bounds on
the RNSNR to support required spectral efficiency of $K$nats/s/Hz.
\begin{theorem} \label{thm:fd}
  For any fixed positive link power gains $Z_{12}$, $Z_{13}$, and
  $Z_{23}$, define the bound
\[
 B_{\mathrm{DF}} \stackrel{\triangle}{=} \left\{
 \begin{array}{ll}
 \displaystyle{\frac{Z_{12}+Z_{23}}{Z_{12}(Z_{13}+Z_{23})}} &
 \mbox{~if~} Z_{12} > Z_{13} \\
 \displaystyle{\frac{1}{Z_{13}}} & \mbox{~otherwise,}
 \end{array} \right.
\]
and
\[
B_{\mathrm{lb}} \stackrel{\triangle}{=}
\displaystyle{\frac{Z_{12}+Z_{13}+Z_{23}}{(Z_{12}+Z_{13})(Z_{13}+Z_{23})}}.
\]
Then $S > B_{\mathrm{DF}}$ is a sufficient condition in order to
support the rate of $K$nats/s/Hz from the source to destination. Also
$S \geq B_{\mathrm{lb}}$ is a necessary condition in order to support
the rate of $K$nats/s/Hz from the source to destination.
\end{theorem}
\begin{proof}
  See Appendix~\ref{app:fd}.
\end{proof}
The RNSNR $B_{\textrm{DF}}$ is achieved by the DF approach employing
the block Markov coding suggested in \cite{Cover79,Cover91}. We also
need to optimally allocate transmit energy between the source and
relay nodes. In addition, the availability of channel state
information (both magnitudes and phases of the fading coefficients of
all three links) as well as symbol timing and carrier phase
synchronization at all the three nodes are implicitly assumed.  The
lower bound $B_{\textrm{lb}}$ is based on the max-flow-min-cut bound
in \cite{Cover91}. No known coding technique can achieve this bound.

\section{Half-Duplex Protocols based on Flow Control}
\label{se:hd}
In this section, we will consider the more practical scenario in which
the relay node operates in the following half-duplex fashion.  We
partition each unit time slot into two sub-slots with respective
durations $t_1$ and $t_2$, where $t_1+t_2=1$. In the first time slot,
the source transmits while the relay and destination receive. In the
second time slot, the source and relay transmit, and the destination
receives.  Based on this half-duplex mode of operation, we will
describe two cooperative communication protocols that make use of the
two basic components of cooperative broadcasting (CB) from the source
to the relay in the first time slot and multiple access (MA) from the
source and relay in the second time slot.  The first protocol does not
require phase synchronization among the three nodes, while the second
protocol does so.

\subsection{Half-Duplex Protocol 1 (HDP1)}
\label{se:p1}
In this protocol, the information from the source to the destination
is divided into three flows of data $x_1$, $x_2$, and $x_3$, where
$x_1+x_2+x_3=K$.  In the first time slot, the source sends, via CB,
two flows of rates $x_1/t_1$ and $x_2/t_1$ to the destination and
relay, respectively.  In the second time slot, the relay and source
send, via MA, two flows of rates $x_2/t_2$ and $x_3/t_2$ to the
destination, respectively.  The information flow of rate $x_2/t_2$
sent by the relay in the second time slot is from the flow of rate
$x_2/t_1$ that it receives and decodes in the first time slot.  We
choose $t_1$, $t_2$, $x_1$, $x_2$, and $x_3$ to minimize the total
power transmitted by the source and relay to support the rate
$K$nats/s/Hz from the source to the destination.

To determine the minimum RNSNR that can support the required rate when
this protocol is employed, we start with the following lemma.
\begin{lemma} \label{thm:cbma}
  \begin{enumerate}
  \item For $0<t_1\leq 1$, the infimum of the SNR required so that the
    source can broadcast at rates $x_1/t_1$ and $x_2/t_1$ to the
    destination and relay, respectively, in the first time slot is
\[ 
S_{\textrm{CB}}= \left \{
\begin{array}{ll}
  \displaystyle
\frac{1}{Z_{12}} ( e^{x_2/t_1} -1) 
 + \frac{1}{Z_{13}} e^{x_2/t_1} ( e^{x_1/t_1} -1) &
\mathrm{~if ~} Z_{13} \geq Z_{12}, \\ 
  \displaystyle
\frac{1}{Z_{13}} ( e^{x_1/t_1} -1) +
\frac{1}{Z_{12}} e^{x_1/t_1} ( e^{x_2/t_1} -1) &
\mathrm{~otherwise}.
\end{array}
\right .
\]
For $t_1=0$, $S_{\textrm{CB}}=0$.
\item For $0<t_2\leq 1$, the infimum of the SNR required so that the
  source and relay can simultaneously transmit at rates $x_3/t_2$ and
  $x_2/t_2$, respectively, to the destination in the second time slot
  is
\[ 
S_{\textrm{MA}}= \left \{
\begin{array}{ll}
  \displaystyle
\frac{1}{Z_{23}} ( e^{x_2/t_2} -1) + \frac{1}{Z_{13}} e^{x_2/t_2} ( e^{x_3/t_2} -1) &
\mathrm{~if ~} Z_{13} \geq Z_{23}, \\ 
  \displaystyle
\frac{1}{Z_{13}} ( e^{x_3/t_2} -1) + \frac{1}{Z_{23}} e^{x_3/t_2} ( e^{x_2/t_2} -1) &
\mathrm{~otherwise}.
\end{array}
\right .
\]
For $t_2=0$, $S_{\textrm{MA}}=0$.
\end{enumerate}
\end{lemma}
\begin{proof}
  See Appendix~\ref{app:cbma}.
\end{proof}

With the help of Lemma~\ref{thm:cbma}, we can now formulate the
optimization of the parameters in Protocol~1 as follows: 
\begin{equation}
\begin{array}{lll}
& \min\ t_1 S_{\textrm{CB}}+ t_2 S_{\textrm{MA}} & \\ 
\mbox{subject~to}
& \mbox{i. total data requirement:} & x_1 + x_2 + x_3 = K \\
& \mbox{ii. total time requirement:} &  t_1+t_2=1 \\
& \mbox{iii. non-negativity requirements:} &  x_1 , x_2 , x_3 , t_1 , t_2 \geq 0  
\end{array}
\label{e:opt}
\end{equation}
where $S_{CB}$ and $S_{MA}$ are of the forms in Lemma~\ref{thm:cbma}.
It is not hard to see that (\ref{e:opt}) is a convex optimization
problem and its solution provides the tightest lower bound for the SNR
required to support the rate of $K$nats/s/Hz:
\begin{theorem} \label{thm:b1}
  Let $B_{1}(K)$ be the minimum value achieved in the optimization
  problem (\ref{e:opt}), normalized by the factor $e^K -1$. Then
  $B_1(K)$ is the infimum of the RNSNR required so that the rate of
  $K$nats/s/Hz can be supported from the source to the destination by
  HDP1.
\end{theorem}

\subsubsection{Description of $B_1(K)$}
\label{se:b1k}
To describe the form of the RNSNR bound $B_1(K)$, we need to consider
the following few cases. This solution is established by applying the
Karush-Kuhn-Tucker (KKT) condition \cite{Boyd04} to the convex
optimization problem (\ref{e:opt}) as detailed in
Appendix~\ref{app:soln}. For notational convenience, we write
\[
M_H(x,y) = \frac{1}{\frac{1}{x}+\frac{1}{y}}
\]
as the harmonic mean\footnote{The definition here actually gives one
  half of the harmonic mean usually defined in the literature. For
  convenience, we will slightly abuse the common terminology and call
  $M_H(x,y)$ the harmonic mean.}  of two real numbers $x$ and $y$.

\paragraph{$Z_{13} \geq M_H(Z_{12},Z_{23})$}
The solution is given by
\[ 
\begin{array}{l}
x_1 = K t_1, \\
x_2 = 0, \\
x_3 = K t_2,
\end{array}
\]
where $t_1$ and $t_2$ can be arbitrarily chosen as long as they
satisfy the non-negativity and total-time requirements. This
corresponds to directly transmitting all data through the link from
the source to destination, without utilizing the relay.  The resulting
value of $B_1(K)$ is
\[
B_1(K) = \frac{1}{Z_{13}}.
\]

\paragraph{$Z_{13} < M_H(Z_{12},Z_{23})$}
Define 
\begin{eqnarray*}
  A_1 &=& Z_{23}  \left( \frac{1}{Z_{13}} - \frac{1}{Z_{12}} \right), 
  \\
  A_2 &=& Z_{12}  \left( \frac{1}{Z_{13}} - \frac{1}{Z_{23}} \right).
\end{eqnarray*}
Notice that $A_1 > 1$ and $A_2 > 1$.  Consider two sub-cases:
\begin{enumerate}
\item[i.] $K> M_H(\log A_1,\log A_2)$: \\
In this case,
\begin{equation}
 B_1(K) = \frac{\min \{ \tilde S_1(K), \tilde S_2(K), \tilde S_3(K)\}}{e^K -1},
\label{e:B1a}
\end{equation}
where the three SNR terms $\tilde S_1(K)$, $\tilde S_2(K)$, and
$\tilde S_3(K)$ are respectively defined in (\ref{e:tS1}),
(\ref{e:tS2}), and (\ref{e:tS3}) below.
  
The first SNR term is given by
\begin{equation}
\tilde S_1(K) =
\min_{\max\left\{0,1-\frac{K}{\log A_1}\right\} \leq t_1 \leq 
  \min\left\{\frac{K}{\log A_2},1\right\}}
  \left\{ \frac{1}{Z_{12}} e^{K + (1-t_{1}) \log A_2} +
\frac{1}{Z_{23}} e^{K + t_{1} \log A_1} - \frac{1}{Z_{13}} \right\}.
\label{e:tS1}
\end{equation}
Define
\[
t^{*} = \frac{\log \left(\displaystyle \frac{Z_{23}\log A_2}{Z_{12}
      \log A_1} \right)+ \log A_2}{\log A_1 + \log A_2}.
\]
Employing the well known inequalities $\log x \leq x-1$ for $x\geq 1$
and $\log x \geq 1 - \frac{1}{x}$ for $x > 0$, it can be shown that $0
\leq t^{*} \leq 1$. By simple calculus, $t^*$ is the minimizing $t_1$
in (\ref{e:tS1}) above when $\max\left\{0,1-\frac{K}{\log A_1}\right\}
\leq t^* \leq \min\left\{\frac{K}{\log A_2},1\right\}$. When $t^*$
lies outside that range, the minimizing $t_1$ must be one of the
boundary points.  When $\tilde S_1(K)$ is the minimum among the three
terms inside the $\min$ operator in (\ref{e:B1a}), the corresponding
solution to the optimization problem (\ref{e:opt}) is given by
\[
\begin{array}{l}
x_1 = K t^{*}- t^{*}(1- t^{*}) \log A_1 , \\
x_2 = t^{*} (1-t^{*}) \log (A_1 A_2) , \\
x_3 = K(1-t^{*})- t^{*} (1-t^{*}) \log A_2, 
\end{array}
\]
with $t_1 = t^*$ and $t_2=1-t^*$.

The second SNR term is given by
\begin{equation}
\tilde S_2(K) = 
\min_{\min\left\{\frac{K}{\log A_2},1\right\} \leq t_{1} \leq 1} \left\{
\frac{t_{1}}{Z_{12}}  e^{K/t_{1}} + \frac{1}{Z_{23}}  e^{K + t_{1}\log A_1 }
- \frac{t_1}{Z_{13}} 
-\frac{1-t_1}{Z_{23}} \right\} .
\label{e:tS2}
\end{equation}
Write the minimizing value of $t_1$ in the expression above as $t^{**}$.
When $\tilde S_2(K)$ is the minimum among the three terms inside the
$\min$ operator in (\ref{e:B1a}), the corresponding solution to
the optimization problem (\ref{e:opt}) is given by
\[
\begin{array}{l}
x_1 = K t^{**}- t^{**} (1-t_{1}) \log A_1 , \\
x_2 = K(1- t^{**}) +t^{**} (1-t^{**}) \log A_1 , \\
x_3 = 0,
\end{array}
\]
with $t_1 = t^{**}$ and $t_2 = 1-t^{**}$.

The third SNR term is given by
\begin{equation}
\tilde S_3(K) = 
\min_{0 \leq t_{1} \leq  \max\left\{0,1- \frac{K}{\log A_1}\right\}}
\left\{ \frac{1-t_{1}}{Z_{23}}  e^{K/(1-t_{1})} 
+ \frac{1}{Z_{12}}  e^{K + (1-t_{1}) \log A_2}   
- \frac{1-t_1}{Z_{13}} -\frac{t_1}{Z_{12}} \right\} .
\label{e:tS3}
\end{equation}
Write the minimizing value of $t_1$ in the expression above as $t^{***}$.
When $\tilde S_3(K)$ is the minimum among the three terms inside the
$\min$ operator in (\ref{e:B1a}), the corresponding solution to
the optimization problem (\ref{e:opt}) is given by
\[
\begin{array}{l}
x_1 = 0, \\
x_2 = K t^{***} +t^{***} (1-t^{***}) \log A_2 , \\
x_3 = K(1- t^{***})- t^{***} (1-t^{***}) \log A_2,
\end{array}
\]
with $t_1 = t^{***}$ and $t_2 = 1-t^{***}$.

\item[ii.] $K \leq M_H(\log A_1, \log A_2)$: \\
In this case,
\begin{equation}
 B_1(K) = \frac{\min \{ \hat S_1(K), \hat S_2(K), \hat S_3(K)\}}{e^K -1},
\label{e:B1b}
\end{equation}
where the three SNR terms $\hat S_1(K)$, $\hat S_2(K)$, and $\hat
S_3(K)$ are respectively defined in (\ref{e:hS1}), (\ref{e:hS2}), and
(\ref{e:hS3}) below.
  
The first SNR term is given by
\begin{equation}
  \hat S_1(K) =
  \min_{ \frac{K}{\log A_2} \leq t_1 \leq 1- \frac{K}{\log A_1}}
  \left\{
    \frac{t_{1}}{Z_{12}}  \left[e^{K/t_{1}}-1\right]
  + \frac{1-t_{1}}{Z_{23}}  \left[e^{K/(1-t_{1})}-1\right] \right\}.
\label{e:hS1}
\end{equation}
Write the minimizing value of $t_1$ in the expression above as $t_{*}$.
When $\hat S_1(K)$ is the minimum among the three terms inside the
$\min$ operator in (\ref{e:B1a}), the corresponding solution to
the optimization problem (\ref{e:opt}) is given by
\[
\begin{array}{l}
x_1=0, \\
x_2=K, \\
x_3=0,
\end{array}
\]
with $t_1 = t_{*}$ and $t_2 = 1-t_{*}$.

The second SNR term is given by
\begin{equation}
\hat S_2(K) =
\min_{1-\frac{K}{\log A_1} \leq t_{1} \leq 1}
\left\{
  \frac{t_{1}}{Z_{12}}  e^{K/t_{1}} + \frac{1}{Z_{23}}  e^{K + t_{1}\log A_1 }  
 - \frac{t_1}{Z_{13}} -\frac{1-t_1}{Z_{23}} \right\} .
\label{e:hS2}
\end{equation}
Write the minimizing value of $t_1$ in the expression above as $t_{**}$.
When $\hat S_2(K)$ is the minimum among the three terms inside the
$\min$ operator in (\ref{e:B1a}), the corresponding solution to
the optimization problem (\ref{e:opt}) is given by
\[ 
\begin{array}{l}
x_1=K t_{**}- t_{**}(1- t_{**}) \log A_1 , \\
x_2=K(1- t_{**}) +t_{**}(1- t_{**}) \log (A_1) , \\
x_3=0,
\end{array}
\] 
with $t_1 = t_{**}$ and $t_2 = 1-t_{**}$.

The third SNR term is given by 
\begin{equation}
\hat S_3(K) = 
\min_{ 0\leq t_{1} \leq \frac{K}{\log A_2}}
\left\{
\frac{1-t_{1}}{Z_{23}}  e^{K/(1-t_{1})} 
+ \frac{1}{Z_{12}}  e^{K + (1-t_{1}) \log A_2}   
- \frac{1-t_1}{Z_{13}} - \frac{t_1}{Z_{12}} \right\}.   
\label{e:hS3}
\end{equation}
Write the minimizing value of $t_1$ in the expression above as $t_{***}$.
When $\hat S_3(K)$ is the minimum among the three terms inside the
$\min$ operator in (\ref{e:B1a}), the corresponding solution to
the optimization problem (\ref{e:opt}) is given by
\[ 
\begin{array}{l}
x_1=0, \\
x_2=K t_{***} +t_{***} (1-t_{***}) \log A_2 , \\
x_3=K(1- t_{***})- t_{***} (1-t_{***}) \log A_2 ,
\end{array}
\] 
with $t_1 = t_{***}$ and $t_2 = 1-t_{***}$.
\end{enumerate}

\subsubsection{Asymptotic-rate scenarios} 
We are interested in characterizing the required RNSNR in the
asymptotic scenarios as the required rate $K$ approaches zero and
infinity, respectively. The following corollary of
Theorem~\ref{thm:b1} and the description of $B_1(K)$ above provides
such characterization:
\begin{corollary} \label{thm:b1asymp}
\begin{enumerate}
\item $B_1(K)$ is continuous and non-decreasing in $K$ for all $K>0$.
\item $\displaystyle \lim_{K \rightarrow 0} B_{1} (K) = \left\{
  \begin{array}{ll}
    \displaystyle
    \frac{1}{Z_{23}} + \frac{1}{Z_{12}} & 
    \mbox{~if~} Z_{13} < M_H(Z_{12},Z_{23})\\
    \displaystyle
    \frac{1}{Z_{13}} &
    \mbox{~if~} Z_{13} \geq M_H(Z_{12},Z_{23}).
  \end{array} \right. $
\item $ \displaystyle
  \lim_{K \rightarrow \infty } B_{1} (K) = \left\{
  \begin{array}{ll}
    \displaystyle
    \frac{A_1^{t^*}}{Z_{23}} + \frac{A_2^{1-t^*}}{Z_{12}} &
    \mbox{~if~} Z_{13} < M_H(Z_{12},Z_{23})\\
    \displaystyle
    \frac{1}{Z_{13}} &
    \mbox{~if~} Z_{13} \geq M_H(Z_{12},Z_{23}).
  \end{array} \right.$
\item $B_1(K)$ is continuous (except at $Z_{13}=Z_{12}=Z_{23}=0$) and
  non-increasing in each of $Z_{13}$, $Z_{12}$ and $Z_{23}$ for all
  $Z_{13}, Z_{12}, Z_{23} \geq 0$.
\end{enumerate}
\end{corollary}
\begin{proof}
  See Appendix~\ref{app:b1asymp}.
\end{proof}

From the solution of the optimization problem described in
Section~\ref{se:b1k} (see the form of solution under (\ref{e:hS1})),
we observe that for a sufficiently low rate requirement, the most
energy-efficient transmission strategy is to select between the direct
link from the source to the destination and the relay path from the
source to the relay and then to the destination. The choice of which
path to take is determined by comparing the power gains of the two
paths. We note that the power gain of the relay path is specified by
the harmonic mean of the power gains of the links from the source to
the relay and from the relay to the destination. The form of
$\lim_{K\rightarrow 0} B_1(K)$ in part 2) of
Corollary~\ref{thm:b1asymp} also suggests this strategy.

When the rate requirement is sufficiently high, the optimal strategy
(see the form of solution under (\ref{e:tS1})) is again to compare the
path gains of the direct and relay paths. If the direct path is
stronger, all information is still sent through this path. Different
from the low-rate case, if the relay path is stronger, most of the
information is still sent through the direct path. Only a fixed amount
(depends on the link power gains, but not on the rate regardless of
how high it is) of information is sent through the relay path. The
reduction of this fixed amount of data through the direct path has the
equivalent effect of improving the fading margin of the direct path and
hence provides diversity advantage. Unlike the low-rate case, this
strategy is not readily revealed by the form of $\lim_{K\rightarrow
  \infty} B_1(K)$ in part 3) of Corollary~\ref{thm:b1asymp}.

\subsection{Half-Duplex Protocol 2 (HDP2)}
\label{se:p2}
In this protocol, the information from the source to the destination
is again divided into three flows of data $x_1$, $x_2$, and $x_3$,
where $x_1+x_2+x_3=K$.  In the first time slot, the source sends, via
CB, two flows of rates $x_1/t_1$ and $x_2/t_1$ to the destination and
relay, respectively, as before.  In the second time slot, the relay
sends the information that it receives in the first time slot to the
destination with a flow of rates $x_2/t_2$. The source, on the other
hand, simultaneously sends two flows of information to the destination
in the second time slot. The first flow is the exact same flow of rate
$x_2/t_2$ sent by the relay. The other flow has rate $x_3/t_2$
containing new information.  Like before, we choose $t_1$, $t_2$,
$x_1$, $x_2$, and $x_3$ to minimize the total power transmitted by the
source and relay to support the rate $K$nats/s/Hz from the source to
the destination.

To send the same flow of data (with rate $x_2/t_2$) in the second time
slot, the source and relay use the same codebook. The codeword symbols
from the source and relay are sent in such a way that the
corresponding received symbols arrive at the destination in phase and
hence add up coherently. In order to do so, the source and relay need
to be phase synchronized and to have perfect channel state information
of the links. We note that these two assumptions are also needed in
the full-duplex approach described in Section~\ref{se:fd}. In
addition, the codebooks used by the source to send the two different
flows in the second time slot are independently selected so that the
transmit power of the source is the sum of the power of the two
codewords sent.

Since the transmission procedure is the same as that of HDP1 in the
first time slot, Lemma~\ref{thm:cbma} part 1) gives the minimum SNR
that can support the required CB transmission in the first time slot.
The minimum SNR required in the second time slot is given by the
following lemma:
\begin{lemma} \label{thm:ipma}
  For $0<t_2\leq 1$, suppose that the source transmits a flow of data
  at rate $x_3/t_2$ to the destination in the second time slot. Then
  the infimum of the SNR required so that the source and relay can
  jointly send another in-phase flow of data at rate $x_2/t_2$ to the
  destination in the second time slot is
\[ 
\hat S_{\mathrm{MA}}= \frac{1}{Z_{13}} ( e^{x_3/t_2} -1) +
\frac{1}{Z_{13}+Z_{23}} e^{x_3/t_2} ( e^{x_2/t_2} -1).
\]
For $t_2=0$, $\hat S_{\mathrm{MA}}=0$.
\end{lemma}
\begin{proof}
  See Appendix~\ref{app:ipma}.
\end{proof}

Let us define $\tilde Z_{23} = Z_{13}+Z_{23}$. Then we note that the
expression of $\hat S_{\mathrm{MA}}$ above can be obtained by putting
$\tilde Z_{23}$ in place of $Z_{23}$ in the expression of
$S_{\mathrm{MA}}$ in Lemma~\ref{thm:cbma}. This means that as far as
minimum SNR is concerned, HDP2 is equivalent to HDP1 with the power
gain of the link from the relay to the destination specified by
$\tilde Z_{23}$ instead.  Using this equivalence, we obtain the
following counterparts of Theorem~\ref{thm:b1} and
Corollary~\ref{thm:b1asymp} for HDP2:
\begin{theorem} \label{thm:b2}
  Let $B_{2}(K)$ be obtained by replacing $Z_{23}$ with $\tilde
  Z_{23}$ in the description of $B_1(K)$ given in
  Section~\ref{se:b1k}. Then $B_2(K)$ is the infimum of the RNSNR
  required so that the rate of $K$nats/s/Hz can be supported from the
  source to the destination by HDP2.
\end{theorem}
We note that $B_2(K) \leq B_1(K)$ since HDP1 can be seen as an
unoptimized version of HDP2 with zero power assigned to the
transmission of the flow of rate $x_2/t_2$ from the source to the
destination during the second time slot.
\begin{corollary} \label{thm:b2asymp}
\begin{enumerate}
\item $B_2(K)$ is continuous and non-decreasing in $K$ for all $K>0$.
\item $\displaystyle \lim_{K \rightarrow 0} B_{2} (K) = \left\{
  \begin{array}{ll}
    \displaystyle
    \frac{1}{\tilde Z_{23}} + \frac{1}{Z_{12}} & 
    \mbox{~if~} Z_{13} < M_H(Z_{12},\tilde Z_{23})\\
    \displaystyle
    \frac{1}{Z_{13}} &
    \mbox{~if~} Z_{13} \geq M_H(Z_{12},\tilde Z_{23}).
  \end{array} \right.$
\item $\displaystyle \lim_{K \rightarrow \infty } B_{2} (K) = \left\{
  \begin{array}{ll}
    \displaystyle
    \frac{\tilde A_1^{\tilde t^*}}{\tilde Z_{23}} +
    \frac{\tilde A_2^{1-\tilde t^*}}{Z_{12}} &
    \mbox{~if~} Z_{13} < M_H(Z_{12},\tilde Z_{23})\\
    \displaystyle
    \frac{1}{Z_{13}} &
    \mbox{~if~} Z_{13} \geq M_H(Z_{12},\tilde Z_{23}).
  \end{array} \right.$
\item $B_2(K)$ is continuous (except at $Z_{13}=Z_{12}=Z_{23}=0$) and
  non-increasing in each of $Z_{13}$, $Z_{12}$ and $Z_{23}$ for all
  $Z_{13}, Z_{12}, Z_{23} \geq 0$.
\end{enumerate}
In parts 2) and 3), $\tilde A_1$, $\tilde A_2$, and $\tilde t^*$ are
the same as $A_1$, $A_2$, and $t^*$, respectively, with $Z_{23}$
replaced by $\tilde Z_{23}$.
\end{corollary}

\section{Performance Analysis}
In this section, we evaluate the performance of HDP1 and HDP2,
particularly in comparison to that of full-duplex relaying.  As
mentioned previously, we model the link power gains $Z_{13}$,
$Z_{12}$, and $Z_{23}$ as i.i.d. exponential random variables with
unit mean. The fading process is assumed to be ergodic and varies
slowly from time slot to time slot.  Hence the minimum RNSNR needed to
support a given rate, or equivalently the maximum achievable rate for
a given RNSNR, is a random variable. Thus we need to consider its
distribution. Moreover, the two protocols, namely HDP1 and HDP2,
considered in Section~\ref{se:hd} are based on the DF approach. The
relay needs to decode in the first time sub-slot and then re-encode to
forward to the destination in the second sub-slot. Hence the decoding
delay is implicitly limited to one\footnote{It is possible for the
  relay to store the signal for a few time slots before decoding, and
  then forward the decoded data to the destination in the next few
  time slots. Nevertheless the decoding delay still needs to be
  finite. We do not consider this time diversity approach here as we
  are primarily interested in the space diversity provided by the
  relay channel.}  time slot. As a result, the maximum ergodic rate
achieved with optimal power control and infinite decoding delay
\cite{Biglieri98} does not apply here. Instead we will consider the
capacity-versus-outage approach of \cite{Ozarow94} (see also
\cite{Biglieri98}) that leads to performance measures like the outage
probability \cite{Ozarow94}, $\varepsilon$-achievable rate
\cite{Caire99}, diversity-multiplexing tradeoff \cite{Zheng03}, and
delay-limited achievable rate \cite{Caire99}.

\subsection{Outage probabilities}
\begin{figure}
  \centering \includegraphics[width=\textwidth]{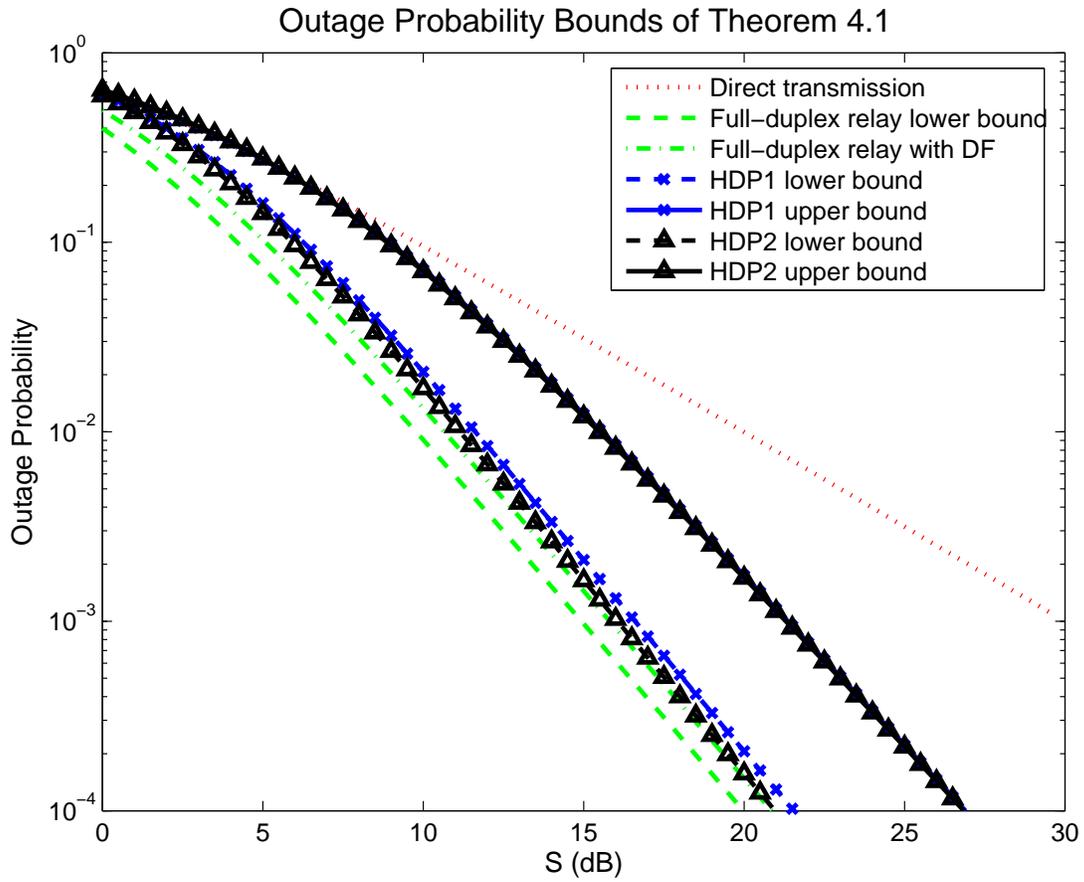}
  \caption{Outage probability bounds of Theorem~\ref{thm:outprob}.}
  \label{f:bound}
\end{figure}

Outage probability is defined as the probability of the event that the
rate $K$ cannot be supported at the RNSNR $S$.  Let us denote the
outage probabilities of full-duplex relaying, full-duplex relaying
with DF, half-duplex relaying using HDP1, and half-duplex relaying
using HDP2 by $P_{\mathrm{fd}}(K,S)$, $P_{\mathrm{DF}}(K,S)$
$P_{1}(K,S)$, and $P_{2}(K,S)$, respectively.  Then by
Theorems~\ref{thm:fd}, \ref{thm:b1}, and \ref{thm:b2}, we have
\begin{eqnarray*}
P_{\mathrm{lb}}(K,S) \stackrel{\triangle}{=}
\Pr (S \leq B_{\mathrm{lb}}) &\leq& P_{\mathrm{fd}}(K,S)
 \leq \Pr (S \leq B_{\mathrm{DF}}) = P_{\mathrm{DF}}(K,S)\\
P_{1}(K,S) & = & \Pr (S \leq B_{1}(K)) \\
P_{2}(K,S) & = & \Pr (S \leq B_{2}(K)).
\end{eqnarray*}
Using these, we can obtain the following bounds on the outage
probabilities. Let $f(x)$ and $g(x)$ be real-valued functions and $a$
be a constant. We say that the function $f(x)$ is of order $ag(x)$
asymptotically, denoted by $f(x) \sim \mathcal{O}(ag(x))$, if
$\lim_{x\rightarrow \infty} f(x)/g(x) = a$. Moreover, we denote the
$\nu$th-order modified Bessel function of the second kind by
$K_\nu(x)$.
\begin{theorem} \label{thm:outprob}
\begin{enumerate}
\item For all $K>0$,
\[
P_{\mathrm{fd}}(K,S) \geq P_{\mathrm{lb}}(K,S) \geq 1 - 2
e^{-\frac{1}{S}} + e^{-\frac{2}{S}} ~\sim
\mathcal{O}\left(\frac{1}{S^2}\right).
\]
\item For all $K>0$,
 \[
 P_{\mathrm{DF}}(K,S) \geq 1 - e^{-\frac{1}{S}} {-\frac{1}{S}}
 e^{-\frac{2}{S}} ~\sim \mathcal{O}\left(\frac{1.5}{S^2}\right).
\]
\item For all $K>0$,
\begin{eqnarray*}
 P_{1}(K,S) &\leq& 
 1 - e^{-\frac{1}{S}} - \int_{0}^{\frac{1}{S}} 2zK_1(2z)e^{-3z}dz \\
&& ~+ \int_{0}^{\frac{\sqrt{2}}{S}} \left[
 2zK_1(2z)e^{-2z} - \frac{4}{S^2 z} 
   K_1\left(\frac{4}{S^2 z}\right) 
   e^{-\frac{4}{S^2 z}} \right] e^{-z} dz 
 ~\sim\mathcal{O}\left(\frac{4\log S}{S^2}\right).
\end{eqnarray*}
\item For all $K>0$,
\[
P_{1}(K,S) \geq \left[1 - \frac{2}{S} K_1\left(\frac{2}{S}\right)
  e^{-\frac{2}{S}} \right] \cdot \left[ 1 - e^{-\frac{1}{S}}\right]
~\sim \mathcal{O}\left(\frac{2}{S^2}\right).
\]
Equality above is achieved when $K$ approaches $0$.
\item For all $K>0$,
\begin{eqnarray*}
 P_{2}(K,S) &\leq& 
 1 - e^{-\frac{1}{S}} - \int_{0}^{\frac{1}{S}} 2zK_1(2z)e^{-2z}dz \\
&& ~+ \int_{0}^{\frac{\sqrt{2}}{S}} \left[
 2zK_1(2z)e^{-2z} - \frac{4}{S^2 z} 
   K_1\left(\frac{4}{S^2 z}\right) 
   e^{-\frac{4}{S^2 z}} \right] dz 
 ~\sim\mathcal{O}\left(\frac{4\log S}{S^2}\right).
\end{eqnarray*}
\item For all $K>0$,
 \[
 P_{2}(K,S) \geq 1 - e^{-\frac{1}{S}} - \frac{2}{S^2}
 K_1\left(\frac{2}{S}\right) e^{-\frac{2}{S}} ~\sim
 \mathcal{O}\left(\frac{1.5}{S^2}\right).
\]
Equality above is achieved when $K$ approaches $0$.
\end{enumerate}
\end{theorem}
\begin{proof}
  See Appendix~\ref{app:outprob}.
\end{proof}
The various bounds in this theorem are illustrated in
Fig.~\ref{f:bound}.

For comparison purpose, it is easy to verify that the outage
probability for direct transmission from the source to destination is
$P_{\mathrm{dt}}(K,S) = \Pr (S \leq 1/Z_{13}) = 1-e^{\frac{1}{S}} \sim
\mathcal{O}\left(\frac{1}{S}\right)$.  From parts 1) and 6) of the
theorem, we see that $\mathcal{O}\left(\frac{1}{S^2}\right) \leq
P_{\mathrm{fd}}(K,S) \leq \mathcal{O}\left(\frac{1.5}{S^2}\right)$.
Hence full-duplex relaying provides a second-order diversity outage
performance as expected. In addition, when the rate requirement is
small and the RNSNR is large, the loss in outage performance due to
the restriction of half-duplex relaying is at most $0.9$dB by using
HDP2.  If phase synchronization between the source and relay is
impractical, then employing HDP1 results in an additional loss of
about $0.6$dB. Comparing parts 2) and 6), we see that HDP2 achieves
the same outage performance as full-duplex relaying based on DF at
asymptotically small rates.  All these observations are readily
illustrated in Fig.~\ref{f:bound}.

\begin{figure}
  \centering \includegraphics[width=\textwidth]{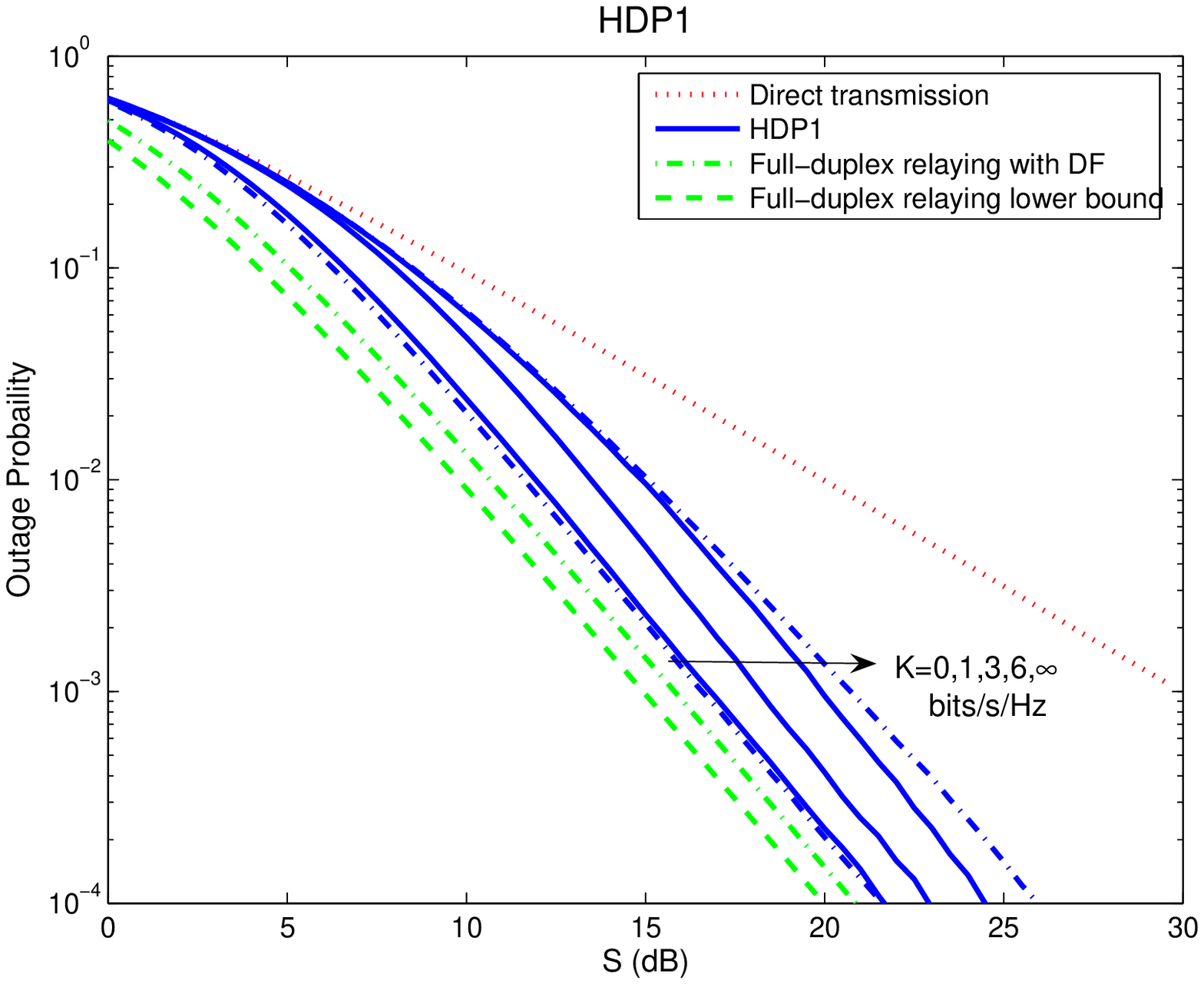}
  \caption{Outage probabilities for HDP1 obtained from Monte Carlo calculations.} \label{f:p1}
\end{figure}

\begin{figure}
  \centering \includegraphics[width=\textwidth]{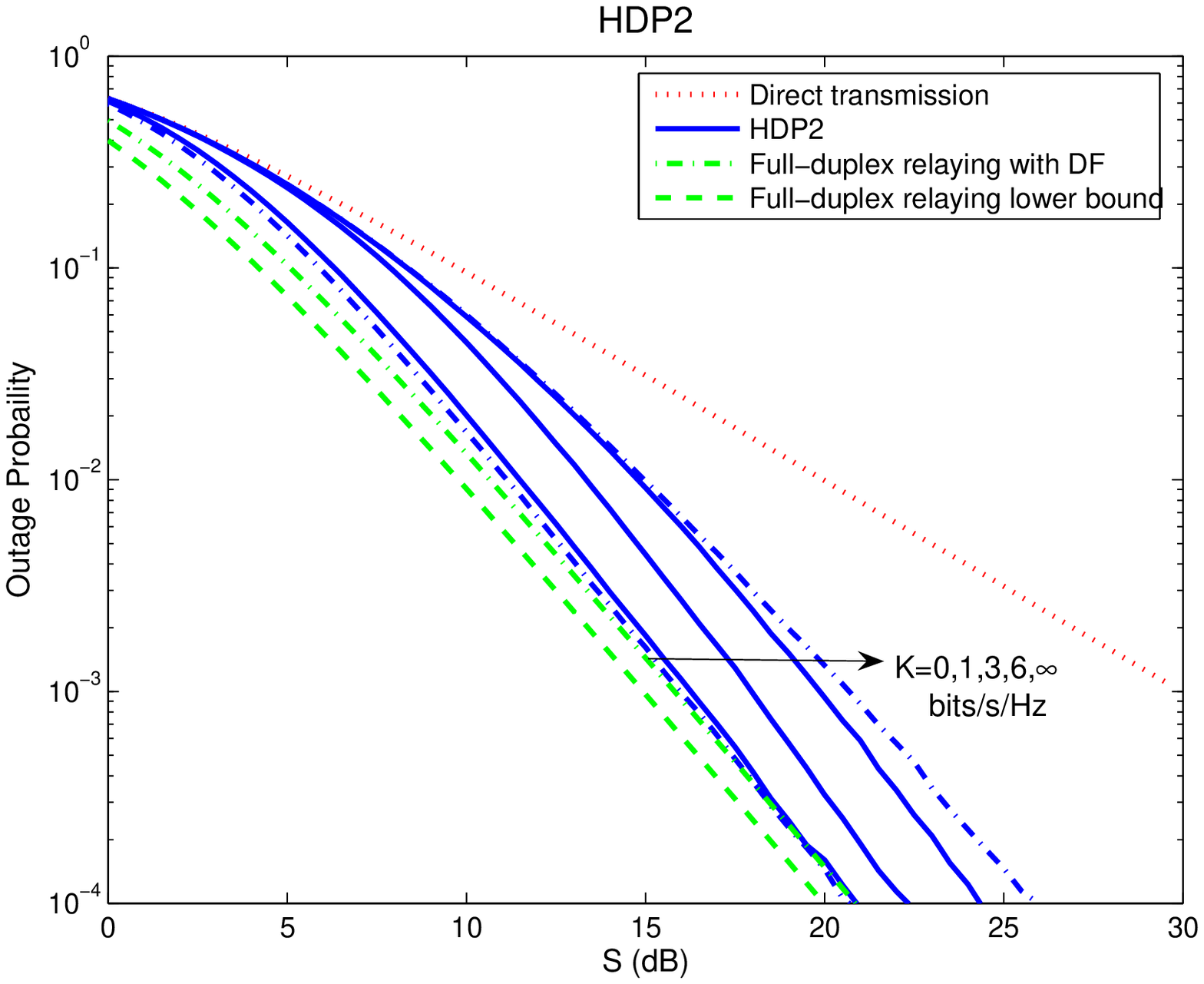}
  \caption{Outage probabilities for HDP2 obtained from Monte Carlo calculations.}  \label{f:p2}
\end{figure}

When the rate requirement increases, the loss of half-duplex relaying
starts to increase. In Figs.~\ref{f:p1} and \ref{f:p2}, we plot the
outage probabilities achieved using HDP1 and HDP2, respectively.  In
each of the figures, we include the outage probabilities when the rate
requirement approaches $0$, $1$, $3$, $6$, and $\infty$ bits/s/Hz. For
comparison, we also plot the lower bounds on outage probabilities for
full-duplex relaying in parts 1) and 2) of Theorem~\ref{thm:fd} and
the outage probability for direction transmission in the figures.  All
the results corresponding to HDP1 and HDP2 in the figures are obtained
using Monte Carlo calculations.  From Fig.~\ref{f:p1}, for HDP1, we
see that the loss, with respect to full-duplex relaying at the outage
probability of $10^{-4}$, is at most $1.5$dB at $1$ bits/s/Hz. The
loss increases to about $2.7$dB and $4.2$dB when the rate increases to
$3$ and $6$ bits/s/Hz, respectively. A similar trend is observed in
Fig.~\ref{f:p2} for HDP2.  At $1$ bits/s/Hz, the loss is about
$0.8$dB.  The loss increases to $2.2$dB and $4.1$dB when the rate
increases to $3$ and $6$ bits/s/Hz, respectively. Moreover, at all the
values of $K$ considered for both HDP1 and HDP2, the simulation
results seem to indicate that the outage probability is of the order
of $\mathcal{O}\left(\frac{a}{S^2}\right)$ for some constant $a$,
whose value is different for the different cases.

When the rate requirement becomes asymptotically large,
Theorem~\ref{thm:outprob} parts 3) and 5) state that the outage
probabilities for HDP1 and HDP2 are at most of order
$\mathcal{O}\left(\frac{4\log S}{S^2}\right)$. This implies that they
both give close-to-second-order diversity performance at
asymptotically high rates. From the simulation results shown in
Figs.~\ref{f:p1} and \ref{f:p2}, it appears that the outage
probabilities for both HDP1 and HDP2 do in fact have the order of
$\mathcal{O}\left(\frac{a\log S}{S^2}\right)$, where $a$ is about
$2.85$. This corresponds to a performance loss of about $5$dB at the
outage probability of $10^{-4}$, and the bound in parts 3) and 5) is
about $0.8$dB loose (cf. Fig.~\ref{f:bound}). We also note that HDP2
does not improve the outage performance, compared to HDP1, at
asymptotically large rates. This is contrary to the finite rate cases
in which HDP2 does provide performance advantage over HDP1, although
the amount of advantage decreases as the rate requirement increases.
In summary, HDP1 seems to be of higher practical utility than HDP2
since the former does not require phase synchronization between the
source and relay, while it only suffers from a performance loss of
about $0.6$dB.

\subsection{$\varepsilon$-achievable rates}
By using the standard sampling representation \cite{Cover91}, the
input-output relationship of the 3-node relay channel over a time slot
can be written as
\begin{eqnarray}
  Y^n &=& \sqrt{Z_{13}} X_1^n + \sqrt{Z_{23}} X_2^n + N^n \nonumber \\
  Y_1^n &=& \sqrt{Z_{12}} X_1^n + N_1^n,
\label{e:model}
\end{eqnarray}
where $X_1^n$, $X_2^n$, $Y^n$, $Y_1^n$, $N^n$, and $N_1^n$ are the
$n$-element transmit symbol vector at the source, transmit symbol
vector at the relay, receive symbol vector at the destination, receive
symbol vector at the relay, Gaussian noise vector at the destination,
and Gaussian noise vector at the relay, respectively. The dimension
$n=2W$ and is assumed to be large. Conditioned on the link gain vector
$Z=[Z_{13}, Z_{12}, Z_{23}]$, the channel is memoryless and described
by the Gaussian conditional pdf $p_{Y^n, Y_1^n| X_1^n, X_2^n,
  Z}(y^n,y_1^n|x_1^n,x_2^n)$.  An $(n,M_n,\varepsilon_n,P_n)$-code
\cite{Han03} over a time slot is one that consists of the encoding and
decoding functions described in \cite{Cover79} allowing one of $M_n$
messages to be sent from the source to destination in a time slot, and
achieves the average (averaged over all codewords sent at the source
and relay, link and noise realizations) error probability of
$\varepsilon_n$, while the maximum (over all codewords) total transmit
energy used in the time slot does not exceed $P_n$.  Since the link
gain vector is available at all nodes, we allow the transmit powers of
the source and relay to vary as functions of the link gains. This
corresponds to the application of power control \cite{Caire99}. As a
result, the power control scheme is included implicitly in the code,
and $P_n$ can be in general a function of the link gain vector $Z$.
In most cases, we are interested in performing power control to
minimize $\varepsilon_n$.  The rate $K$ is
$(\varepsilon,P_t)$-achievable if there exists a sequence of
$(n,M_n,\varepsilon_n,P_n)$-codes satisfying $\limsup_{n\rightarrow
  \infty} \varepsilon_n \leq \varepsilon$, $\liminf_{n \rightarrow
  \infty} \frac{1}{n}\log M_n \geq K$, and $\limsup_{n\rightarrow
  \infty} P_n \leq P_t$ almost surely (a.s.).

For half-duplex relaying, when the relay listens (transmits), its
transmit (receive) symbols are restricted to zero. In the previous
sections, we have assumed that the relay first listens for $t_1$
seconds in a time slot and then transmits in the remaining $t_2$
seconds.  In this case, it is more convenient to describe the channel
by the CB and MA conditional pdfs, $p_{Y^{t_1 n}, Y_1^{t_1 n}|
  X_1^{t_1 n},Z}(y^{t_1n},y_1^{t_1 n}|x_1^{t_1 n})$ and $p_{Y^{t_2 n}|
  X_1^{t_2 n} X_2^{t_2 n},Z}(y^{t_2n}|x_1^{t_2 n},x_2^{t_2 n})$, for
the first and second sub-slots, respectively. We will also say that
the rate $K$ is $(\varepsilon,P_t)$-achievable with HDP1 (HDP2) if
there exists a sequence of $(n,M_n,\varepsilon_n,P_n)$-codes with the
CB and MA coding in the first and second sub-slots as described in
HDP1 (HDP2), satisfying $\limsup_{n\rightarrow \infty} \varepsilon_n
\leq \varepsilon$, $\liminf_{n \rightarrow \infty} \frac{1}{n}\log M_n
\geq K$, and $\limsup_{n\rightarrow \infty} P_n \leq P_t$ a.s. 

The theorem below states that the various $\varepsilon$-achievable
rates are characterized by the corresponding outage probabilities
defined in the previous section.
\begin{theorem}
\label{thm:ec}
\begin{enumerate}
\item For all $\varepsilon > 0$ and $0 < \delta < 1$, if the rate $K$
  is $(\varepsilon,P_t)$-achievable, then \\
  $\varepsilon \geq \delta P_{\mathrm{lb}} \left(K,
    \frac{e^K-1}{e^{(1-\delta) K} - 1} S \right)$.

\item For all $\varepsilon > 0$, the rate $K$ is
  $(\varepsilon,P_t)$-achievable if $P_\mathrm{DF}(K,S) \leq
  \varepsilon$.
  
\item For all $\varepsilon > 0$, the rate $K$ is
  $(\varepsilon,P_t)$-achievable with HDP1 if $P_{1}(K,S) \leq 
  \varepsilon$.
  
\item For all $\varepsilon > 0$, the rate $K$ is
  $(\varepsilon,P_t)$-achievable with HDP2 if $P_{2}(K,S) \leq
  \varepsilon$.
\end{enumerate}
\end{theorem}
\begin{proof}
  See Appendix~\ref{app:ec}.
\end{proof}

\subsection{Diversity-multiplexing tradeoff}
It is also interesting to investigate the diversity-multiplexing
tradeoff of \cite{Zheng03} for HDP1 and HDP2. To this end, we need to
follow \cite{Laneman04} to change the parameterization of the outage
probabilities from $(K,S)$ to $(\tilde K, \tilde S)$, where $\tilde S$
is the SNR and $\tilde K$ is the multiplexing gain ($0 < \tilde K <
1$) defined by
\[
\tilde K = \frac{K}{\log (1+\tilde S)}.
\]
With the parameterization $(\tilde K, \tilde S)$, the diversity orders
\cite{Zheng03} achieved by HDP1 and HDP2 are defined as
\[
\Delta_i(\tilde K) = \lim_{\tilde S \rightarrow \infty} \frac{-\log
  P^i_e (\tilde K,\tilde S)}{\log \tilde S},
\]
where $P^i_e(\tilde K,\tilde S)$ is the average error probability of
HDP$i$ at SNR $\tilde S$ and multiplexing gain $\tilde K$, for $i=1$
and $2$, respectively. Then the diversity orders can be readily
obtained in the following corollary of Theorems~\ref{thm:outprob} and
\ref{thm:ec}.
\begin{corollary} \label{thm:dm}
  For $i=1$ and $2$, $\Delta_i(\tilde K) = 2(1-\tilde K)$. Hence HDP1
  and HDP2 achieve the maximum diversity advantage possible for the
  relay channel when link gain information is available at all nodes.
\end{corollary}
\begin{proof}
  First, by Theorem~\ref{thm:ec}, $P^i_e(\tilde K,\tilde S) \leq
  P_i(\tilde K,\tilde S)$ for $i=1$ and $2$. Also notice that since
  $\tilde S = S \left(e^K -1\right)$, $\displaystyle S = \frac{\tilde
    S}{(1+\tilde S)^{\tilde K} - 1}$.  Hence $S \sim
  \mathcal{O}\left(\tilde S^{1 - \tilde K}\right)$.  As a result,
  applying parts 3) -- 6) of Theorem~\ref{thm:outprob} with the
  parameterization $(\tilde K, \tilde S)$, for sufficiently large
  $\tilde S$ and $i=1,2$,
\[
\mathcal{O}\left(\frac{a_i}{\tilde S^{2(1-\tilde K)}}\right) \leq
P_i(\tilde K,\tilde S) \leq 
\mathcal{O}\left(\frac{4(1-\tilde K)\log \tilde S}{\tilde S^{2(1-\tilde K)}}
\right),
\]
where $a_1 = 2$ and $a_2 = 1.5$. Applying $-\log$, dividing the result
by $\log \tilde S$, and finally taking limit as $\tilde S \rightarrow
\infty$ on each item in the inequality equation above give
$\Delta_i(\tilde K) \geq 2(1-\tilde K)$ for $i=1$ and $2$.
On the other hand, part 1) of Theorem~\ref{thm:outprob} and part 1) of
Theorem~\ref{thm:ec} force the error probability of any transmission
scheme over the relay channel to be larger than $\delta
\mathcal{O}\left(\frac{e^{1-\delta}}{\tilde S^{2(1-\tilde K)}}\right)$
for all $0 < \delta < 1$. As a result, the maximum possible diversity
order of any transmission scheme over the relay channel is $2(1-\tilde
K)$. Thus we have the desired result.
\end{proof}

\subsection{Delay-limited rates}
When the average error probability decreases to zero, the
$\varepsilon$-achievable rates becomes the delay-limited rates
\cite{Caire99}. We calculate the delay-limited rates achievable by
HDP1 and HDP2 in this section.

We first employ the following definition \cite{Han03} as our
definition of delay limited rate: The rate $K$ is $P_t$-achievable if
there exists a sequence of $(n,M_n,\varepsilon_n,P_n)$-codes over a
time slot, satisfying $\lim_{n\rightarrow \infty} \varepsilon_n =0$,
$\liminf_{n \rightarrow \infty} \frac{1}{n}\log M_n \geq K$, and
$\limsup_{n\rightarrow \infty} P_n \leq P_t$ a.s. .  Unfortunately,
part 1) of Theorem~\ref{thm:ec} forces the $P_t$-achievable rate of
any transmission scheme over the relay channel to be zero as long as
$P_t$ is finite. This is due to the Rayleigh fading nature of the
links and the restriction that the total transmit energy in each time
slot needs to be bounded by $P_t$. It turns out that more meaningful
results can be obtained if we relax the latter restriction.

Recall that the link power gains vary independently from time slot to
time slot. With power control to maintain the error probability
$\varepsilon_n$, the total transmit energy $P_n$ (a function of $Z$)
of an $(n,M_n,\varepsilon_n,P_n)$-code may vary from time slot to time
slot.  This may require the total transmit energy to be very large in
the worst faded time slots.  As a relaxation of the transmit energy
constraint, we require the average total transmit energy per time slot
over many time slots to be bounded. Then the ergodicity of the fading
process requires $E[P_n]$ to be bounded. This relaxation motivates the
following definition: The rate $K$ is \emph{long-term}
$P_t$-achievable if there exists a sequence of
$(n,M_n,\varepsilon_n,P_n)$-codes over a time slot, satisfying
$\lim_{n\rightarrow \infty} \varepsilon_n =0$, $\liminf_{n \rightarrow
  \infty} \frac{1}{n}\log M_n \geq K$, and $\limsup_{n\rightarrow
  \infty} E[P_n] \leq P_t$.

The following theorem then specifies the delay-limited rates
achievable by HDP1 and HDP2:
\begin{theorem} \label{thm:dlc}
  For $K>0$, define 
\begin{eqnarray*}
  P^{\mathrm{lb}}_t(K) &=& E[B_{\mathrm{lb}}] (e^K - 1) N_0 W \\
  P^{\mathrm{DF}}_t(K) &=& E[B_{\mathrm{DF}}] (e^K - 1) N_0 W \\
  P^{1}_t(K) &=& E[B_{1}(K)] (e^K - 1) N_0 W \\
  P^{2}_t(K) &=& E[B_{2}(K)] (e^K - 1) N_0 W.
\end{eqnarray*}  
Then $ P^{\mathrm{lb}}_t(K) \leq P^{\mathrm{DF}}_t(K) \leq P^{2}_t(K)
\leq P^{1}_t(K) < \infty$.
\begin{enumerate}
\item If the rate $K$ is long-term $P_t$-achievable, then $P_t
  \geq P^{\mathrm{lb}}_t(K)$.
\item The rate $K$ is long-term $P^{\mathrm{DF}}_t(K)$-achievable.
\item The rate $K$ is long-term $P^{1}_t(K)$-achievable with HDP1.
\item The rate $K$ is long-term $P^{2}_t(K)$-achievable with HDP2.
\end{enumerate}
\end{theorem}
\begin{proof}
  See Appendix~\ref{app:dlc}.
\end{proof}

In Fig.~\ref{f:dlrate}, we plot the rate $K$ against
$\frac{P_t^{i}(K)}{N_0W}$ for $i \in \{\mathrm{lb},\mathrm{DF}\}$. For
$i \in \{1,2\}$, notice that 
\[
E[B_i(0)](e^K-1) \leq \frac{P_t^{i}(K)}{N_0W} \leq
E[B_i(\infty)](e^K-1)
\]
by Corollaries~\ref{thm:b1asymp} and \ref{thm:b2asymp}, respectively.
In each case, we plot the lower and upper bounds instead. The true
$\frac{P_t^{i}(K)}{N_0W}$ curve lies between the bounding curves. Also
the true curve approaches the lower bound when $K$ is small and the
upper bound when $K$ is large. For comparison, we also plot the curve
$\frac{e^K-1}{N_0W}$ which corresponds to the SNR required to achieve
the rate $K$ in an additive white Gaussian noise (AWGN) channel with
unit power gain. Thus, at each rate, the loss of performance, with
respect to an AWGN channel, in dB for approach $i$ due to link fading
and the restriction of one-slot decoding delay is
$E[B_i]_{\mathrm{dB}} = 10\log_{10}(E[B_i])$. The results obtained
from numerical calculations are shown in Table~\ref{tb:loss}. From the
table, we see that the loss when employing full-duplex relaying is
between $2.17$dB and $2.76$dB, where the upper limit on the loss can
be achieved by DF with optimal power control. The loss when using HDP1
with optimal power control ranges from $3.33$dB to $5.45$dB, while the
loss when using HDP2 with optimal power control is between $3.02$dB
and $5.36$dB. The loss of performance of half-duplex relaying with
respect to full-duplex relaying is at most $3.28$dB. This loss happens
when the rate requirement is very large and HDP1 is employed. When the
rate is very small, the loss drops down to at most $0.85$dB with the
use of HDP2.  With the delay-limited rate as performance measure, HDP1
once again appears to be a good tradeoff between complexity and
performance. The maximum loss when using HDP1 instead of HDP2 is only
$0.31$dB.

\section{Conclusions}
With channel state information available at all nodes, we have shown
that a half-duplex cooperative transmission design, based on
optimizing distinct flows through the direct link from the source to
the destination and the path via the relay, can effectively harness
diversity advantage of the relay channel in both high-rate and
low-rate scenarios. Specifically, the proposed design gives outage
performance approaching that of full-duplex relaying using
decode-and-forward at asymptotically low rates. When the rate
requirement becomes asymptotically large, the design still gives a
close-to-second-order outage diversity performance.
The design also gives the best diversity-multiplexing tradeoff
possible for the relay channel. With optimal long-term power control
over the fading relay channel, the proposed design can give
delay-limited rate performance that is within a few dBs of the
capacity performance of the additive white Gaussian channel in both
low-rate and high-rate scenarios.

In addition to the good performance, a perhaps more important
advantage of the proposed relaying design is that only flow-level
design is needed to optimize the use of the rather standard components
of cooperative broadcasting and multiple access.  This advantage makes
generalizations of the design to more-complicated relay networks
manageable. In general, the availability of channel information at the
nodes appears to simplify cooperative transmission designs. Thus it is
worthwhile to investigate how to effectively spread the channel state
information in a wireless network.

\appendix

\subsection{Proof of Theorem~\ref{thm:fd}}
\label{app:fd}
First consider the sufficient condition in 
Theorem~\ref{thm:fd}. We consider two different cases:

\subsubsection{$Z_{12} > Z_{13}$}
Let $0 \leq \alpha \leq 1$ be the fraction of the total energy, $P_t$,
allocated to the source node.  Then the transmit energy of the relay
node is $(1-\alpha)P_t$.  By \cite[Theorem~1]{Cover79} (also see
\cite{Host05}), the following rate is achievable by the relay channel
when the relay decodes and re-encodes its received signal:
\begin{eqnarray}
R_{\textrm{DF}} (\alpha) &=& \max_{0\leq \beta \leq 1} \min\left\{
  C\left( \frac{P_t}{N_0 W} \left[ \alpha Z_{13} + (1-\alpha) Z_{23} + 2 
      \sqrt{(1-\beta)\alpha (1-\alpha) Z_{13}Z_{23}}\right]\right), \right.
\nonumber \\
 && 
 ~~~~~~~~~~~~~~~
  \left. C\left( \frac{P_t}{N_0 W} \alpha \beta Z_{12} \right) \right\},
  \label{e:Rfdu}
\end{eqnarray}
where $C(x) = \log (1+x)$. We can further maximize the rate by
optimally allocating transmit energy between the source and relay,
i.e.,
\begin{eqnarray*}
R_{\textrm{DF}} &=& \max_{0\leq\alpha\leq 1} R_{\textrm{DF}} (\alpha) \\
&=& 
C\left( \frac{P_t}{N_0 W} \max_{0\leq \alpha, \beta \leq 1} \min\left\{
  \alpha Z_{13} + (1-\alpha) Z_{23} + 2 
      \sqrt{(1-\beta)\alpha (1-\alpha) Z_{13}Z_{23}},
  \alpha \beta Z_{12} \right\} \right),
\end{eqnarray*}
where the second equality results from the fact that $C(x)$ is an
increasing function. Hence, the requirement that the RNSNR satisfies
$S > 1/Z_{\textrm{DF}}$ is sufficient for $R_{\textrm{DF}} = K$, where
\begin{equation}
Z_{\textrm{DF}} = 
\max_{0\leq \alpha, \beta \leq 1} \min\left\{
  \alpha Z_{13} + (1-\alpha) Z_{23} + 2 
      \sqrt{(1-\beta)\alpha (1-\alpha) Z_{13}Z_{23}},
  \alpha \beta Z_{12} \right\}.
\label{e:Zfd}
\end{equation}
Thus it reduces to solving the optimization problem in
(\ref{e:Zfd}).

To solve (\ref{e:Zfd}), we write $\displaystyle Z_{\textrm{DF}}
(\alpha)= \max_{0\leq \beta \leq 1} \min\left\{ \alpha Z_{13} +
  (1-\alpha) Z_{23} + 2 \sqrt{(1-\beta)\alpha (1-\alpha)
    Z_{13}Z_{23}}, \alpha \beta Z_{12} \right\}.$ and consider two
cases:
\paragraph{$\displaystyle 0 \leq \alpha \leq \frac{Z_{23}}{Z_{23}+Z_{12}-Z_{13}}$}
Under this case, the second term inside the $\min$ operator is smaller
than the first term for all $0 \leq \beta \leq 1$. Hence
$Z_{\textrm{DF}} (\alpha) = \max_{0\leq \beta \leq 1} \alpha \beta
Z_{12} = \alpha Z_{12}$.
\paragraph{$\displaystyle \frac{Z_{23}}{Z_{23}+Z_{12}-Z_{13}} < \alpha \leq 1$} 
Under this case, notice that the first and second terms inside the
$\min$ operator are strictly decreasing and increasing in $\beta$,
respectively. Moreover the two terms equalize at some $0 \leq \beta_*
\leq 1$. Hence $Z_{\textrm{DF}} (\alpha) = \alpha \beta_* Z_{12}$.
Solving for the equalizing $\beta_*$, we get
\[
Z_{\textrm{DF}} (\alpha) = \alpha Z_{13} + (1-\alpha) Z_{23} \left( 1-
  2\frac{Z_{13}}{Z_{12}}\right) + 2\sqrt{(1-\alpha)
  \frac{Z_{13}}{Z_{12}} \left(1-\frac{Z_{13}}{Z_{12}}\right) Z_{23}
  \left[\alpha Z_{12} - (1-\alpha) Z_{23} \right]}.
\]

\noindent Now we maximize $Z_{\textrm{DF}} (\alpha)$ over $0 \leq \alpha
\leq 1$.  For case a), $\displaystyle \max_{\alpha} Z_{\textrm{DF}}
(\alpha) = \frac{Z_{12}Z_{23}}{Z_{23}+Z_{12}-Z_{13}}$. For case b), a
direct but tedious calculation shows that $\displaystyle \max_{\alpha}
Z_{\textrm{DF}} (\alpha) =
\frac{Z_{12}(Z_{13}+Z_{23})}{Z_{12}+Z_{23}}$. It is not hard to verify
that the maximum value in case b) is larger than the maximum value in
case a). Hence $\displaystyle Z_{\textrm{DF}} =
\frac{Z_{12}(Z_{13}+Z_{23})}{Z_{12}+Z_{23}}$, and the sufficient
condition is $\displaystyle S >
\frac{Z_{12}+Z_{23}}{Z_{12}(Z_{13}+Z_{23})}$.

\subsubsection{$Z_{12} \leq Z_{13}$}
First note that the capacity of the relay channel is upper bounded by
the maximum sum rate of the CB channel from the source to the relay
and destination. This CB channel is a degraded Gaussian broadcast
channel, and the individual rates $R_{13}$ and $R_{12}$ from the
source to the destination and relay, respectively, satisfy \cite[Ch.
14]{Cover91}
\begin{eqnarray}
R_{13} &<& C\left(\frac{\alpha Z_{13} P_t}{N_0 W} \right) \nonumber \\
R_{12} &<& C\left(\frac{(1-\alpha) Z_{12}P_t}{\alpha Z_{12} P_t + N_0 W}\right),
\label{e:cbregion}
\end{eqnarray}
for any $0 \leq \alpha \leq 1$. To have $R_{13}+R_{12} \geq K$, we need
\begin{eqnarray*}
K &<& \max_{0 \leq \alpha \leq 1}
\left\{C\left(\frac{\alpha Z_{13} P_t}{N_0 W} \right) +
  C\left(\frac{(1-\alpha) Z_{12}P_t}{\alpha Z_{12} P_t + N_0
    W}\right) \right\} \\
&=&  \log\left[\left(1+Z_{12}\frac{P_t}{N_0 W}\right) \cdot
 \max_{0 \leq \alpha \leq 1}
  \left(
  \frac{1+\alpha Z_{13} \frac{P_t}{N_0 W}}{1+\alpha Z_{12} \frac{P_t}{N_0 W}}
\right) \right]\\
&=& 
  C\left(Z_{13} \frac{P_t}{N_0 W}\right),
\end{eqnarray*}
where the last equality is obtained by choosing $\alpha =1$, due to
the condition that $Z_{13} \geq Z_{12}$. Hence $S > 1/Z_{13}$.  This
lower bound corresponds to sending all information directly form the
source to the destination without using the relay.

For the necessary condition, we employ the max-flow-min-cut bound of
\cite[Theorem~14.10.1]{Cover91} to obtain an upper bound,
$R_{\textrm{lb}} (\alpha)$, on the rate of the relay channel. It turns
out \cite{Host05} that the expression for $R_{\textrm{lb}} (\alpha)$
is obtained simply by replacing every occurrence of $Z_{12}$ by
$Z_{12}+Z_{13}$ in (\ref{e:Rfdu}) above. In addition, the power
optimization procedure in case 1) above carries through directly for
this case with every occurrence of $Z_{12}$ replaced by
$Z_{12}+Z_{13}$. Thus we obtain the necessary condition as
$\displaystyle S \geq
\frac{Z_{12}+Z_{13}+Z_{23}}{(Z_{12}+Z_{13})(Z_{13}+Z_{23})}$.

\subsection{Proof of Lemma~\ref{thm:cbma}}
\label{app:cbma}
\begin{enumerate}
\item The case of $t_1=0$ trivially requires $x_1=x_2=0$, and hence
  $S_{\textrm{CB}} =0$. So we consider $0 < t_1 \leq 1$. If
  $Z_{13}>Z_{12}$, we have a degraded broadcast channel during this
  time slot. Thus rate constraints in (\ref{e:cbregion}) must be
  satisfied with $R_{13}=x_1/t_1$ and $R_{12} = x_2/t_1$. Combining
  the two inequalities to remove $\alpha$, it is easy to obtain the
  stated lower bound $S_{\textrm{CB}}$ of the SNR $P_t/N_0 W$. We note
  that this lower bound corresponds to the optimal choice $\alpha =
  \frac{\displaystyle ( e^{x_1/t_1} -1)/ {Z_{13}}}{\displaystyle (
    e^{x_2/t_1} -1)/Z_{12} + e^{x_2/t_1} ( e^{x_1/t_1} -1)/Z_{13}}$.
  Interchanging the roles of $Z_{13}$ and $Z_{12}$, we get the stated
  SNR lower bound for the case of $Z_{13} \leq Z_{12}$.
  
\item The case of $t_2=0$ trivially requires $x_2=x_3=0$, and hence
  $S_{\textrm{MA}} =0$. So we consider $0 < t_2 \leq 1$. The capacity
  region of this Gaussian MA channel is specified by \cite[Ch.
  14]{Cover91}:
  \begin{eqnarray}
  \frac{x_3}{t_2} &<& C\left(\frac{\alpha Z_{13}P_t}{N_0 W} \right),
  \nonumber \\
  \frac{x_2}{t_2} &<& C\left(\frac{(1-\alpha) Z_{23}P_t}{N_0 W} \right),
  \nonumber \\
  \frac{x_2+x_3}{t_2} &<& C\left(\frac{[\alpha Z_{13} + (1-\alpha) Z_{23}] P_t}{N_0 W} \right),
  \label{e:maregion}
  \end{eqnarray}
  for any $0 \leq \alpha \leq 1$, where $\alpha$ and $1-\alpha$ are
  the fractions of the transmit power assigned to the source and
  relay, respectively.  We want to optimally choose $\alpha$ so that
  the SNR $P_t/N_0 W$ required to satisfy (\ref{e:maregion}) is
  minimized. First, suppose that $Z_{13} > Z_{23}$. Then rearranging
  the second and third inequalities in (\ref{e:maregion}) gives
  \begin{eqnarray*}
  \alpha &<& 1 - \frac{1}{Z_{23}} \left(e^{x_2/t_2} - 1\right)
  \frac{N_0 W}{P_t}, \\  
  \alpha &>& \frac{1}{Z_{13}-Z_{23}} \cdot \left\{ \left[
      e^{(x_2+x_3)/t_2} -1\right] \frac{N_0 W}{P_t} - Z_{23} \right\},
  \end{eqnarray*}
  respectively. Combining these two inequalities to remove $\alpha$,
  we obtain the stated lower bound $S_{\textrm{MA}}$. We note that the
  corresponding optimal choice $\alpha=\frac{\displaystyle (
    e^{x_3/t_2} -1)/ {Z_{13}}}{\displaystyle ( e^{x_2/t_2} -1)/Z_{23}
    + e^{x_2/t_2} ( e^{x_3/t_2} -1)/Z_{13}}$.
  
  Interchanging the roles of $Z_{13}$ and $Z_{23}$, we get the stated
  SNR lower bound for the case of $Z_{13} < Z_{23}$. When
  $Z_{13}=Z_{23}$, the third inequality in (\ref{e:maregion}) gives
  the stated lower bound $S_{\textrm{MA}}$. The choice of optimal
  $\alpha$ in this case is exactly the same as the one in the case of
  $Z_{13}>Z_{23}$ (or $Z_{13}<Z_{23}$).
\end{enumerate}

\subsection{Solution to optimization problem (\ref{e:opt})}
\label{app:soln}
Suppose that $t_1$ and $t_2$ are fixed, satisfying both the
non-negativity and total-time requirements.  Then we can view the
optimization problem (\ref{e:opt}) as a convex optimization problem in
$x_1$, $x_2$ and $x_3$.  Rewriting it in standard form \cite{Boyd04}:
\begin{equation}
\begin{array}{lll}
& 
\displaystyle \min_{x_1,x_2,x_3} \tilde S(t_1,t_2) =  t_1 S_{\mathrm{CB}}+ t_2 S_{\mathrm{MA}} & \\ 
\mbox{Subject to} 
& \mbox{i. total-data requirement:} & x_1 + x_2 + x_3 = K \\
& \mbox{ii. non-negativity requirements:} &  -x_1 , -x_2 , -x_3 \leq 0   
\end{array}
\label{e:optsim}
\end{equation}
Since this optimization problem is convex, the rate tuple
$(x_1,x_2,x_3)$ is a solution if it satisfies the following KKT
conditions:
\begin{enumerate}
\item[K1.] $\displaystyle \nabla \tilde S + \sum_{i=1}^{3} \lambda_i
  \nabla (- x_i) + \mu \nabla (x_1+x_2+x_3-K) = 0$

\item[K2.] $\lambda_i (-x_i) = 0$ for $ i =1, 2, 3$ 

\item[K3.] $\lambda_i \geq 0$ for $i =1, 2, 3$ 

\item[K4.] $-x_i \leq 0$ for $i =1, 2, 3$

\item[K5.] $x_1 + x_2 + x_3 = K$.
\end{enumerate} 

Our approach to solve the original optimization problem (\ref{e:opt})
is to first solve the sub-problem (\ref{e:optsim}) for each pair of
$(t_1,t_2)$, and then minimize $\min_{x_1,x_2,x_3} \tilde S(t_1,t_2)$
over all allowable pairs. To this end, we consider the following cases
and obtain solution to the optimization problem (\ref{e:optsim}) by
directly checking the KKT conditions. Note that we assume in below
that both $t_1$ and $t_2$ are positive. For $t_2=0$ ($t_1=0$),
Lemma~\ref{thm:cbma} tells us that the transmission in the first
(second) sub-slots reduces trivially to transmission over the direct
link from the source to the destination. Hence
$\min_{x_1,x_2,x_3}\tilde S(1,0)=\min_{x_1,x_2,x_3}\tilde
S(0,1)=\left(e^K-1\right)/Z_{13}$.

\subsubsection{$Z_{13} \geq Z_{12}$ and $Z_{13} \geq Z_{23}$}
From Lemma~\ref{thm:cbma},
\[ 
\tilde S(t_1,t_2) = \frac{t_1}{Z_{12}} \left( e^{x_2/t_1} -1 \right) +
\frac{t_1}{Z_{13}} e^{x_2/t_1} \left( e^{x_1/t_1} -1 \right) +
\frac{t_2}{Z_{23}} \left( e^{x_2/t_2} -1 \right) + \frac{t_2}{Z_{13}}
e^{x_2/t_2} \left( e^{x_3/t_2} -1 \right).
\]
The condition K1 yields
\[
\begin{array}{l}
\displaystyle
\frac{1}{Z_{13}} e^{(x_1+x_2)/t_1} - \lambda_1 + \mu = 0 \\
\displaystyle
\frac{1}{Z_{12}} e^{x_2/t_1} + \frac{1}{Z_{13}} e^{x_2/t_1} ( e^{x_1/t_1} -1) + 
\frac{1}{Z_{23}} e^{x_2/t_2} + \frac{1}{Z_{13}} e^{x_2/t_2} ( e^{x_3/t_2} -1) -
\lambda_2 + \mu = 0   \\
\displaystyle
\frac{1}{Z_{13}} e^{(x_2+x_3)/t_2} - \lambda_3 + \mu = 0.
\end{array}
\]
It is then easy to check that the following solution satisfy the KKT conditions:
\[
\begin{array}{ll}
x_1=K t_1, & \lambda_1 = 0 \\
\displaystyle
x_2=0, & \displaystyle \lambda_2 = \frac{1}{Z_{12}} + \frac{1}{Z_{23}} - \frac{1}{Z_{13}} + \frac{1}{Z_{13}} ( e^{K} -1)  
\\
x_3=K t_2, & \lambda_3 = 0 \\
\displaystyle
\mu = - \frac{1}{Z_{13}} e^{K}.
\end{array}
\]

\subsubsection{$Z_{13} \geq Z_{12}$ and $Z_{13} < Z_{23}$} 
From Lemma~\ref{thm:cbma},
\[
\tilde S(t_1,t_2) = \frac{t_1}{Z_{12}} \left( e^{x_2/t_1} -1\right) +
\frac{t_1}{Z_{13}} e^{x_2/t_1} \left( e^{x_1/t_1} -1\right) +
\frac{t_2}{Z_{13}} \left( e^{x_3/t_2} -1\right) + \frac{t_2}{Z_{23}}
e^{x_3/t_2} \left( e^{x_2/t_2} -1\right).
\]
The condition K1 yields
\[ 
\begin{array}{l}
\displaystyle
\frac{1}{Z_{13}} e^{(x_1+x_2)/t_1} - \lambda_1 + \mu = 0 \\
\displaystyle
\frac{1}{Z_{12}} e^{x_2/t_1} + \frac{1}{Z_{13}} e^{x_2/t_1} ( e^{x_1/t_1} -1)  
+ \frac{1}{Z_{23}} e^{(x_2+x_3)/t_2}  -\lambda_2 + \mu = 0   \\
\displaystyle
\frac{1}{Z_{13}} e^{x_3/t_2} +  \frac{1}{Z_{23}} e^{x_3/t_2} ( e^{x_2/t_2} -1) - \lambda_3 + \mu = 0.
\end{array}
\]
It is then easy to check that the following solution satisfy the KKT conditions:
\[
\begin{array}{ll}
x_1=K t_1, & \lambda_1 = 0 \\
\displaystyle
x_2=0, & \displaystyle \lambda_2 = \frac{1}{Z_{12}} - \frac{1}{Z_{13}} +  \frac{1}{Z_{23}} e^{K}  
\\
x_3=K t_2, & \lambda_3 = 0 \\
\displaystyle
\mu = - \frac{1}{Z_{13}} e^{K} .
\end{array}
\]

\subsubsection{$Z_{13} < Z_{12}$ and $Z_{13} \geq Z_{23}$} 
From Lemma~\ref{thm:cbma},
\[
\tilde S(t_1,t_2) = \frac{t_1}{Z_{13}} \left( e^{x_1/t_1} -1\right) +
\frac{t_1}{Z_{12}} e^{x_1/t_1} \left( e^{x_2/t_1} -1\right) +
\frac{t_2}{Z_{23}} \left( e^{x_2/t_2} -1\right) + \frac{t_2}{Z_{13}}
e^{x_2/t_2} \left( e^{x_3/t_2} -1\right).
\]
The condition K1 yields
\[ 
\begin{array}{l}
\displaystyle
\frac{1}{Z_{13}} e^{x_1/t_1} + \frac{1}{Z_{12}} e^{x_1/t_1} ( e^{x_2/t_1} -1)  - \lambda_1 + \mu = 0 \\
\displaystyle
\frac{1}{Z_{12}} e^{(x_1+x_2)/t_1}  + 
\frac{1}{Z_{23}} e^{x_2/t_2} + \frac{1}{Z_{13}} e^{x_2/t_2} ( e^{x_3/t_2} -1) -\lambda_2 + \mu = 0   \\
\displaystyle
\frac{1}{Z_{13}} e^{(x_2+x_3)/t_2} - \lambda_3 + \mu = 0.
\end{array}
\]
It is then easy to check that the following solution satisfy the KKT conditions:
\[
\begin{array}{ll}
x_1=K t_1, & \lambda_1 = 0 \\
x_2=0, & \displaystyle \lambda_2 =
\frac{1}{Z_{12}} e^{K} + \frac{1}{Z_{23}}  - \frac{1}{Z_{13}} 
\\
x_3=K t_2, & \lambda_3 = 0 \\
\displaystyle
\mu = - \frac{1}{Z_{13}} e^{K} .
\end{array}
\]

\subsubsection{$M_H(Z_{12},Z_{23}) \leq Z_{13} < \min\{Z_{12},Z_{23}\}$} 
From Lemma~\ref{thm:cbma},
\[
\tilde S(t_1,t_2)=\frac{t_1}{Z_{13}} \left( e^{x_1/t_1} -1\right) +
\frac{t_1}{Z_{12}} e^{x_1/t_1} \left( e^{x_2/t_1} -1\right) +
\frac{t_2}{Z_{13}} \left( e^{x_3/t_2} -1\right) + \frac{t_2}{Z_{23}}
e^{x_3/t_2} \left( e^{x_2/t_2} -1\right).
\]
The condition K1 yields
\[ 
\begin{array}{l}
\displaystyle
\frac{1}{Z_{13}}e^{x_1/t_1} + \frac{1}{Z_{12}} e^{x_1/t_1} ( e^{x_2/t_1} -1) - \lambda_1 + \mu = 0 \\
\displaystyle
\frac{1}{Z_{12}} e^{(x_1+x_2)/t_1}  
+ \frac{1}{Z_{23}} e^{(x_2+x_3)/t_2}  -\lambda_2 + \mu = 0   \\
\displaystyle
\frac{1}{Z_{13}} e^{x_3/t_2} + \frac{1}{Z_{23}} e^{x_3/t_2} (  e^{x_2/t_2}-1 ) - \lambda_3 + \mu = 0.
\end{array}
\]
It is then easy to check that the following solution satisfy the KKT
conditions:
\[
\begin{array}{ll}
x_1=K t_1, & \lambda_1 = 0 \\
x_2=0, & 
\displaystyle \lambda_2 = (\frac{1}{Z_{12}} + \frac{1}{Z_{23}} - \frac{1}{Z_{13}}) e^{K} 
\\
x_3=K t_2, & \lambda_3 = 0 \\
\displaystyle
\mu = - \frac{1}{Z_{13}} e^{K} .
\end{array}
\]

\subsubsection{${Z_{13}} < M_H({Z_{12}},{Z_{23}})$}
The expression for $\tilde S(t_1,t_2)$ in case 4) still holds.
However, we need to consider the following two sub-cases in order to
express the solution to the optimization problem (\ref{e:optsim}):

\paragraph{ $K> M_H(\log A_1,\log A_2)$}
\begin{enumerate}
\item[i.] For $1- \frac{K}{\log A_1} \leq t_1 \leq \frac{K}{\log
    A_2}$, the following solution satisfies the KKT conditions:
\[
\begin{array}{ll}
x_1=K t_1- t_1 t_2 \log A_1, & \lambda_1 = 0 \\
x_2=t_1 t_2 \log (A_1 A_2), & \lambda_2 = 0 \\
x_3=K t_2 - t_1 t_2 \log A_2 , & \lambda_3 = 0 \\
\displaystyle
\mu = -\frac{1}{Z_{12}} e^{K+t_2 \log A_2} -\frac{1}{Z_{23}} e^{K+t_1 \log A_1}.
\end{array}
\]

\item[ii.]  For $\frac{K}{\log A_2} < t_1 < 1$, the following solution
  satisfies the KKT conditions:
\[
\begin{array}{ll}
x_1=K t_1- t_1 t_2 \log A_1, & \lambda_1 = 0 \\
x_2=K t_2 + t_1 t_2 \log A_1 , & \lambda_2 = 0 \\
x_3=0, & 
\displaystyle
\lambda_3 = \frac{1}{Z_{13}} - \frac{1}{Z_{23}} - \frac{1}{Z_{12}} e^{K/t_1}\\
\displaystyle
\mu = - \frac{1}{Z_{23}} e^{ K + t_1 \log A_1 } - \frac{1}{Z_{12}} e^{K/t_1}.
\end{array}
\]

\item[iii.] For $0 < t_1 < 1-\frac{K}{\log A_1}$,  
the following solution satisfies the KKT conditions:
\[
\begin{array}{ll}
x_1=0, & 
\displaystyle
\lambda_1 = \frac{1}{Z_{13}} - \frac{1}{Z_{12}} -  \frac{1}{Z_{23}} e^{K/t_2} \\
x_2=K t_1 + t_1 t_2 \log A_2 , & \lambda_2 = 0 \\
x_3=K t_2- t_1 t_2 \log A_2 , & \lambda_3 = 0 \\
\displaystyle
\mu = - \frac{1}{Z_{12}} e^{ K + t_2 \log A_2 } - \frac{1}{Z_{23}} e^{K/t_2}.
\end{array}
\]
\end{enumerate}

\paragraph{$K \leq M_H(\log A_1,\log A_2)$}
\begin{enumerate}
\item[i.] For $\frac{K}{\log A_2} \leq t_1 \leq 1- \frac{K}{\log
    A_1}$, the following solution satisfies the KKT conditions:
\[
\begin{array}{ll}
x_1=0, &
\displaystyle
\lambda_1 = \frac{1}{Z_{13}} - \frac{1}{Z_{12}} - \frac{1}{Z_{23}} e^{K/t_2} \\
x_2=K,& \lambda_2 = 0
\\
x_3=0, &
\displaystyle
\lambda_3 = \frac{1}{Z_{13}} - \frac{1}{Z_{23}} - \frac{1}{Z_{12}} e^{K/t_1} \\
\displaystyle
\mu = - \frac{1}{Z_{12}} e^{K/t_1} -  \frac{1}{Z_{23}} e^{K/t_2} .
\end{array}
\]

\item[ii.] For $1- \frac{K}{\log A_1} < t_1 < 1$, the following
  solution satisfies the KKT conditions:
\[
\begin{array}{ll}
x_1=K t_1- t_1 t_2 \log A_1, & \lambda_1 = 0 \\
x_2=K t_2 + t_1 t_2 \log A_1 , & \lambda_2 = 0 \\
x_3=0, &
\displaystyle
\lambda_3 = \frac{1}{Z_{13}} - \frac{1}{Z_{23}} -  \frac{1}{Z_{12}} e^{K/t_1} \\
\displaystyle
\mu = - \frac{1}{Z_{23}} e^{ K + t_1 \log A_1 } - \frac{1}{Z_{12}} e^{K/t_1}.
\end{array}
\]

\item[iii.] For $0 < t_1 < \frac{K}{\log A_2}$, the following solution
  satisfies the KKT conditions:
\[
\begin{array}{ll}
x_1=0, &
\displaystyle
\lambda_1 = \frac{1}{Z_{13}} - \frac{1}{Z_{12}} - \frac{1}{Z_{23}} e^{K/t_2} \\
x_2=K t_1 + t_1 t_2 \log A_2 , & \lambda_2 = 0
\\
x_3=K t_2- t_1 t_2 \log A_2 , & \lambda_3 = 0 \\
\displaystyle
\mu = - \frac{1}{Z_{12}} e^{ K + t_2 \log A_2 } - \frac{1}{Z_{23}} e^{K/t_2}.
\end{array}
\]
\end{enumerate}

For cases 1)--4), direction substitution of the solution yields
$\min_{x_1,x_2,x_3}\tilde S(t_1,t_2)=\left(e^K-1\right)/Z_{13}$. Since
this solution is independent of the choice of $(t_1,t_2)$, the
solution to the optimization problem (\ref{e:opt}) in these four cases
is simply $\left(e^K-1\right)/Z_{13}$. Also note that these four cases
can be collectively specified by the condition $Z_{13} \geq
M_H(Z_{12},Z_{23})$.

For case 5a), the three functions $\tilde S_1(K)$, $\tilde S_2(K)$,
and $\tilde S_3(K)$ respectively described in (\ref{e:tS1}),
(\ref{e:tS2}), and (\ref{e:tS3}) can be obtained by direct
substitution of the solutions in the 3 sub-cases (i., ii., and iii,
respectively), and then minimizing the corresponding
$\min_{x_1,x_2,x_3} \tilde S(t_1,t_2)$ over the range of $t_1$
specified in each sub-cases. Hence the final solution of the
optimization problem (\ref{e:opt}) is obtained by finding the minimum
among the these three functions. For case 5b), a similar procedure
yields the fact that the solution to the optimization problem
(\ref{e:opt}) is the minimum among the three functions $\hat S_1(K)$,
$\hat S_2(K)$, and $\hat S_3(K)$ respectively described in
(\ref{e:hS1}), (\ref{e:hS2}), and (\ref{e:hS3}).

\subsection{Proof of Corollary~\ref{thm:b1asymp}}
\label{app:b1asymp}
The proof of the results in this corollary is based on the fact that
$B_1(K)$ is the (normalized) solution to the optimization problem
(\ref{e:opt}) and the form of $B_1(K)$ described in
Section~\ref{se:b1k}.

\begin{enumerate}
\item Fix $Z_{13}$, $Z_{12}$, and $Z_{23}$. From the description of
  $B_1(K)$ in Section~\ref{se:b1k}, since $B_1(K)$ is trivially
  continuous and non-decreasing in $K$ when under the condition
  $Z_{13} \geq M_H(Z_{12},Z_{23})$, it suffices to consider $B_1(K)$
  under the condition $Z_{13} < M_H(Z_{12},Z_{23})$, which is assumed
  for the rest of the proof.
  
  For convenience, let us denote the respective functions inside the
  $\min$ operators of $\tilde S_1(K)$, $\tilde S_2(K)$, $\tilde
  S_3(K)$, $\hat S_1(K)$, $\hat S_2(K)$, and $\hat S_3(K)$ by the
  addition of the sign $'$. Define, for
  $0 \leq t_1 \leq 1$,
  \begin{equation}
  \tilde B(K,t_1) = \left\{
  \begin{array}{ll}
    \frac{\tilde S_1'(K)}{e^K-1} & \mbox{~if~} 
    K > M_H(\log A_1, \log A_2) \mbox{~and~}
    \max\left\{0,1-\frac{K}{\log A_1}\right\} \leq t_1
    \leq \min\left\{\frac{K}{\log A_2},1\right\}
    \\
    \frac{\tilde S_2'(K)}{e^K-1} & \mbox{~if~} 
    K > M_H(\log A_1, \log A_2) \mbox{~and~}
    \min\left\{\frac{K}{\log A_2},1\right\} \leq t_1 \leq 1
    \\
    \frac{\tilde S_3'(K)}{e^K-1} & \mbox{~if~} 
    K > M_H(\log A_1, \log A_2) \mbox{~and~}
    0 \leq t_1 \leq \max\left\{0,1-\frac{K}{\log A_1}\right\}
    \\
    \frac{\hat S_1'(K)}{e^K-1} & \mbox{~if~} 
    0< K \leq M_H(\log A_1, \log A_2) \mbox{~and~}
    \frac{K}{\log A_2} \leq t_1 \leq 1 - \frac{K}{\log A_1}
    \\
    \frac{\hat S_2'(K)}{e^K-1} & \mbox{~if~} 
    0< K \leq M_H(\log A_1, \log A_2) \mbox{~and~}
    1-\frac{K}{\log A_1} \leq t_1 \leq 1
    \\
    \frac{\hat S_3'(K)}{e^K-1} & \mbox{~if~} 
    0< K \leq M_H(\log A_1, \log A_2) \mbox{~and~}
    0 \leq t_1 \leq \frac{K}{\log A_2}.
  \end{array} \right.
  \label{e:tB}
  \end{equation}
  Notice that for any fixed $0 \leq t_1 \leq 1$, $\tilde B(K,t_1)$ is
  piecewise continuous, with six pieces over the respective ranges of
  $K$ that they are defined. Also it is easy to check that at each end
  point where two adjacent pieces meet, the values of the pieces
  coincide (and hence the definition of $\tilde B(K,t_1)$ above is
  valid). Thus $\tilde B(K,t_1)$ is continuous in $K$. The same
  argument with fixed $K$ shows that $\tilde B(K,t_1)$ is continuous
  in $t_1$. Then $B_1(K) = \displaystyle \min_{0\leq t_1 \leq 1}
  \tilde B(K,t_1)$ and hence is continuous in $K$.

  Now to show $B_1(K)$ is non-decreasing in $K$, it suffices to show
  that, for each fixed $0\leq t_1 \leq 1$, the functions $\frac{\tilde
    S_1'(K)}{e^K -1}$, $\frac{\tilde S_2'(K)}{e^K -1}$, $\frac{\tilde
    S_3'(K)}{e^K -1}$, $\frac{\hat S_1'(K)}{e^K -1}$, $\frac{\hat
    S_2'(K)}{e^K -1}$, and $\frac{\hat S_3'(K)}{e^K -1}$ are all
  non-decreasing in the corresponding ranges of $K$ that the functions
  are used in the definition of $\tilde B(K,t_1)$ in (\ref{e:tB})
  above. To this end, we will repeatedly employ the following form of
  Young's inequality:
  \[
  x^t y^{1-t} \leq tx + (1-t)y,
  \]
  for nonnegative $x$,$y$, and $0\leq t\leq1$.
  
  Fix $0\leq t_1 \leq 1$. First let us consider $\tilde S_1(K)$. For
  $K$ in the corresponding range in (\ref{e:tB}),
  \[
  \frac{d~}{dK} \frac{\tilde S_1'(K)}{e^K-1} = \left(\frac{1}{Z_{13}}
    - \frac{A_1^{t^*}}{Z_{23}} - \frac{A_2^{1-t^*}}{Z_{12}}\right)
  \frac{e^K}{\left(e^K-1\right)^2}.
  \]
  But by Young's inequality,
  \begin{eqnarray*}
  \frac{1}{Z_{13}} - 
  \frac{A_1^{t^*}}{Z_{23}} - \frac{A_2^{1-t^*}}{Z_{12}}
  &=& 
  \frac{1}{Z_{13}} -
  \left(\frac{1}{Z_{23}}\right)^{1-t^*}
  \left(\frac{1}{Z_{13}} - \frac{1}{Z_{12}} \right)^{t^*} -
  \left(\frac{1}{Z_{12}}\right)^{t^*}
  \left(\frac{1}{Z_{13}} - \frac{1}{Z_{23}} \right)^{1-t^*} \\
  &\geq& 
  \frac{1}{Z_{13}} -
  \frac{1-t^*}{Z_{23}} - \frac{t^*}{Z_{13}} + \frac{t^*}{Z_{12}} -
  \frac{t^*}{Z_{12}} - 
  \frac{1-t^*}{Z_{13}} + \frac{1-t^*}{Z_{23}} \\
  &=& 0.
  \end{eqnarray*}
  Thus $\frac{\tilde S_1'(K)}{e^K -1}$ is non-decreasing.
  
  Next consider $\tilde S_2'(K)$.  For $K$ in the corresponding range
  in (\ref{e:tB}),
  \[
  \frac{d~}{dK} \frac{\tilde S_2'(K)}{e^K-1} = \left\{\frac{1}{Z_{12}}
    \left[ (1-t) e^{K/t} +t - e^{K(1-t)/t} \right] +
    \frac{1-t-A_1^t}{Z_{23}} - \frac{t}{Z_{12}} + \frac{t}{Z_{13}}
  \right\} \cdot \frac{e^K}{\left(e^K-1\right)^2}.
  \]
  Again by Young's inequality,
  \[
  (1-t) e^{K/t} +t - e^{K(1-t)/t} \geq e^{K(1-t)/t} \cdot 1^t -
  e^{K(1-t)/t} =0
  \]
  and
  \begin{eqnarray*}
  \frac{1-t-A_1^t}{Z_{23}} - \frac{t}{Z_{12}} + \frac{t}{Z_{13}} &=&
  \frac{1-t}{Z_{23}} - \frac{t}{Z_{12}} + \frac{t}{Z_{13}} -
  \left(\frac{1}{Z_{23}}\right)^{1-t} 
  \left(\frac{1}{Z_{13}} - \frac{1}{Z_{12}} \right)^{t} \\
  &\geq&
  \frac{1-t}{Z_{23}} - \frac{t}{Z_{12}} + \frac{t}{Z_{13}} -
  \frac{1-t}{Z_{23}} - \frac{t}{Z_{13}} + \frac{t}{Z_{12}} \\
  &=& 0.
  \end{eqnarray*}
  Hence $\frac{\tilde S_2'(K)}{e^K -1}$ is non-decreasing. Finally, we
  note that the non-decreasing nature of the functions $\frac{\tilde
    S_3'(K)}{e^K -1}$, $\frac{\hat S_1'(K)}{e^K -1}$, $\frac{\hat
    S_2'(K)}{e^K -1}$, and $\frac{\hat S_3'(K)}{e^K -1}$ can be proven
  in the same way.

\item When $K$ is sufficiently small, $\displaystyle B_1(K) =
  \frac{\hat S_1(K)}{e^K -1}$. Then a simple application of
  L'Hospital's rule gives the desired result.

\item When $K$ is sufficiently large, $\displaystyle B_1(K) =
  \frac{\tilde S_1(K)}{e^K -1}$. Then simply taking limit gives the
  desired result.
  
\item Fix $K$ and $t_1$. Augment the definition of $\tilde B(K,t_1)$
  in (\ref{e:tB}) by adding $\tilde B(K,t_1) = \frac{1}{Z_{13}}
  \frac{1}{e^K-1}$ when $Z_{13} \geq M_H(Z_{12},Z_{23})$.  For the
  rest of the proof, this augmented $\tilde B(K,t_1)$ will be
  considered as a function of $Z_{13}$, $Z_{12}$, and $Z_{23}$,
  despite its notation.  Then an argument similar to the one in part
  1) can be employed to show that $B_1(K)$ is continuous in each of
  $Z_{13}$, $Z_{12}$, and $Z_{23}$ for every $K > 0$, except for
  $Z_{13}=Z_{12}=Z_{23}=0$ at which $B_1(K)$ becomes infinite.
  
  Again similar to part 1), in order to show $B_1(K)$ is
  non-increasing in each of $Z_{13}$, $Z_{12}$, and $Z_{23}$, we only
  need to show that for each fixed $K$ and $t_1$, the functions
  $\tilde S_1'(K)$, $\tilde S_2'(K)$, $\tilde S_3'(K)$, $\hat
  S_1'(K)$, $\hat S_2'(K)$, and $\hat S_3'(K)$ are all non-increasing
  in each of $Z_{13}$, $Z_{12}$, and $Z_{23}$, over the respective
  ranges of these functions shown in (\ref{e:tB}). Indeed, this fact
  can be shown by verifying that the derivatives involved are all
  non-positive. The only interesting case is $\frac{d \tilde
    S_1'(K)}{d Z_{13}}$, which needs the use of Young's inequality:
  \begin{eqnarray*}
  \frac{d\tilde S_1'(K)}{dZ_{13}} &=&
  -\frac{1}{Z_{13}^2} \left\{ e^K \left[t_1 A_1^{t_1-1} +(1-t_1) A_2^{-t_1}
    \right] -1 \right\} \\
  & \leq & 
  -\frac{1}{Z_{13}^2} \left\{ e^K \left(A_1 A_2\right)^{-t_1(1-t_1)}
  -1\right\} \\
  & = & 
  -\frac{1}{Z_{13}^2} \left\{ e^{K-t_1(1-t_1)(\log A_1 + \log A_2)}
  -1\right\} \\
  & \leq & 
  -\frac{1}{Z_{13}^2} \left\{ e^{K-t_1 K - (1-t_1) K}
  -1\right\} = 0,
  \end{eqnarray*}
  where the second line is due to Young's inequality and the last line
  is due to the fact that $K \geq (1-t_1)\log A_1$ and $K \geq t_1\log
  A_2$ in the range of interest of $\tilde S_1'(K)$.
\end{enumerate}

\subsection{Proof of Lemma~\ref{thm:ipma}}
\label{app:ipma}

The case of $t_2=0$ trivially requires $x_2=x_3=0$, and hence $\hat
S_{\textrm{MA}} =0$. So we consider $0 < t_2 \leq 1$. Suppose that the
transmit power of the relay is $P_2$ and the transmit power of the
source is $P_1+P_3$, where $P_1$ is the power employed to transmit the
flow of rate $x_2/t_2$ while $P_3$ is the power of the flow of rate
$x_3/t_2$.  Then the transmission procedure in the second time slot of
HDP2 (cf. Section~\ref{se:p2}) describes the transmission over an
equivalent two-user Gaussian MA channel in which one user of rate
$x_2/t_2$ has power
$\left(\sqrt{Z_{13}P_{1}}+\sqrt{Z_{23}P_{2}}\right)^2$ and another
user of rate $x_3/t_2$ has power $Z_{13}P_3$.  From the capacity
region of this Gaussian MA channel specified by \cite[Ch.
14]{Cover91}, $P_1$, $P_2$ and $P_3$ must satisfy:
\begin{eqnarray*}
  \frac{x_3}{t_2} & < & C\left( \frac{Z_{13} P_{3}}{N_0 W} \right)
  \nonumber \\
  \frac{x_2}{t_2} & < &
  C\left( \frac{(\sqrt{Z_{13}P_{1}}+\sqrt{Z_{23}P_{2}})^2}{N_0 W} \right)
  \nonumber \\
  \frac{x_2+x_3}{t_2} & < &
  C\left(\frac{(\sqrt{Z_{13}P_{1}}+\sqrt{Z_{23}P_{2}})^2+Z_{13}P_{3}}{N_0 W}
  \right).
  \label{e:ipmaregion}
\end{eqnarray*}
To minimize the total power (energy) given the rates of transmission,
we consider the following optimization problem:
\[
\begin{array}{ll}
& \min P_1 + P_2 + P_3 \\ 
\mbox{subject~to}
& f_1({\bf P}) = a - Z_{13} P_{3} \leq 0 \\
& f_2({\bf P}) = b - (\sqrt{Z_{13}P_{1}}+\sqrt{Z_{23}P_{2}})^2 \leq 0 \\
& f_3({\bf P}) = c - (\sqrt{Z_{13}P_{1}}+\sqrt{Z_{23}P_{2}})^2 - Z_{13}P_{3} \leq 0 \\
& f_4({\bf P}) = -P_1 \leq 0 \\
& f_5({\bf P}) = -P_2 \leq 0 \\
& f_6({\bf P}) = -P_3 \leq 0 ,
\end{array}
\]
where ${\bf P}=(P_1, P_2, P_3)$, $a = N_0 W (e^{x_3/t_2} -1)$, 
$b = N_0 W (e^{x_2/t_2} -1)$, and $c= N_0 W (e^{(x_2+x_3)/t_2} -1)$. 
Notice that $c \geq a + b$. 
It can be shown that this is a convex optimization problem.  The power tuple
$(P_1,P_2,P_3)$ is a solution if it satisfies the following KKT
conditions:
\begin{enumerate}
\item[K1.] $\displaystyle \nabla (P_1+P_2+P_3) + \sum_{i=1}^{6} \lambda_i
  \nabla f_i({\bf P}) = 0$

\item[K2.] $\lambda_i f_i({\bf P})= 0$ for $ i =1, 2, \ldots, 6$ 

\item[K3.] $\lambda_i \geq 0$ for $i =1, 2, \ldots, 6$ 

\item[K4.] $f_i ({\bf P}) \leq 0$ for $i =1, 2, \ldots, 6$.
\end{enumerate} 
The condition K1 yields
\[ 
\begin{array}{l}
\displaystyle
1- (\lambda_2 + \lambda_3) (\sqrt{Z_{13}P_{1}}+\sqrt{Z_{23}P_{2}}) \sqrt{\frac{Z_{13}}{P_1}} - \lambda_4 =0
\\
\displaystyle
1- (\lambda_2 + \lambda_3) (\sqrt{Z_{13}P_{1}}+\sqrt{Z_{23}P_{2}}) \sqrt{\frac{Z_{23}}{P_2}} - \lambda_5 =0
\\
\displaystyle
1 - (\lambda_1+\lambda_3) Z_{13} - \lambda_6 =0 .
\end{array}
\]
It is then easy to check that the following solution satisfies the KKT conditions:
\[
\begin{array}{ll}
\displaystyle
P_1= \frac{(c-a)Z_{13}}{(Z_{13}+Z_{23})^2}, &
\displaystyle
\lambda_1= \frac{1}{Z_{13}} - \frac{1}{Z_{13}+Z_{23}}, \\
\displaystyle
P_2= \frac{(c-a)Z_{23}}{(Z_{13}+Z_{23})^2}, &
\displaystyle
\lambda_3= \frac{1}{Z_{13}+Z_{23}},\\
\displaystyle
P_3= \frac{a}{Z_{13}}, &
\displaystyle
\lambda_2=\lambda_4=\lambda_5=\lambda_6=0 .
\end{array}
\]
Then normalizing the sum of this choice of $P_1$, $P_2$, and $P_3$ by
$N_0 W$ gives the stated expression of $\hat S_{\textrm{MA}}$ in
Lemma~\ref{thm:ipma}.

\subsection{Proof of Theorem~\ref{thm:outprob}}
\label{app:outprob}
To prove the theorem, we need to use the following result:
\begin{claim} \label{c1}
For any $x\geq z \geq 0$,
\[
\Pr \left( \frac{1}{Z_{12}} + \frac{1}{Z_{23}+z} \geq \frac{1}{x} \right)
= 1 - 2xK_1(2x)e^{-2x+z}.
\]
\end{claim}
\begin{proof}
\begin{eqnarray*}
\Pr \left( \frac{1}{Z_{12}} + \frac{1}{Z_{23}+z} \geq \frac{1}{x} \right)
&=& 
\int_{0}^{\infty} \Pr \left( \frac{1}{Z_{12}} \geq  
  \frac{1}{x} - \frac{1}{y+z} \right) e^{-y} dy \\
&=& 
\int_{0}^{x-z} e^{-y} dy + \int_{x-z}^{\infty} \left( 1 -
  e^{-\frac{1}{\frac{1}{x}-\frac{1}{y+z}}} \right) e^{-y} dy \\
&=& 
 1 - \int_{x-z}^{\infty} 
  e^{-\left(\frac{1}{\frac{1}{x}-\frac{1}{y+z}} +y \right)} dy \\
&=& 
 1 - e^{-2x+z} \int_{0}^{\infty} e^{-\left(y+\frac{x^2}{y}\right)} dy,
\end{eqnarray*}
where the integral in the last line is an integral representation of
the function $2xK_1(2x)$ \cite[pp. 53]{Tranter68} or \cite[pp.
969]{Grad94}.
\end{proof}
We note that the same result is obtained for the special case of $z=0$
in \cite{Hasna03} using moment generating functions of exponential
random variables.

\begin{enumerate}
\item By Theorem~\ref{thm:fd},
\begin{eqnarray*}
P_{\mathrm{fd}}(K,S)  \geq P_{\mathrm{lb}} (K,S) & = & \Pr \left(S \leq
  \frac{Z_{12}+Z_{13}+Z_{23}}{(Z_{12}+Z_{13})(Z_{13}+Z_{23})}\right) \\
& \geq & 
\Pr \left( \left\{ S \leq \frac{1}{Z_{12}+Z_{13}} \right\}
  \cup \left\{ S \leq \frac{1}{Z_{13}+Z_{23}}\right\} \right) \\
&=&
  1 - 
  \Pr \left( \left\{ S > \frac{1}{Z_{12}+Z_{13}} \right\}
  \cap \left\{ S > \frac{1}{Z_{13}+Z_{23}}\right\} \right) \\
&=&
  \int_{0}^{\infty} \left[ 1 - \Pr\left( Z_{12} >
        \frac{1}{S}-z\right) \cdot \Pr\left( Z_{23} >
          \frac{1}{S} - z\right) \right] e^{-z} dz \\
&=&
  \int_{0}^{\infty} e^{-z} dz -
  \int_{\frac{1}{S}}^{\infty} e^{-z} dz - 
  \int_{0}^{\frac{1}{S}} e^{-\frac{2}{S} +z} dz \\
&=&
  1 - 2  e^{-\frac{1}{S}} + e^{-\frac{2}{S}}.
\end{eqnarray*}
It is also easy to see that $\lim_{S\rightarrow \infty} \frac{ 1 - 2
  e^{-\frac{1}{S}} + e^{-\frac{2}{S}}}{1/S^2} = 1$.

\item By Theorem~\ref{thm:fd},
\begin{eqnarray*}
P_{\mathrm{DF}}(K,S) & = & \Pr (S \leq B_{\mathrm{DF}})\\
&=&
\underbrace{\Pr
  \left(S \leq \frac{1}{Z_{13}} \Big| Z_{13} \geq Z_{12}\right) \cdot
  \Pr (Z_{13} \geq Z_{12})}_{a} \\
& & + ~\underbrace{\Pr
  \left(S \leq \frac{Z_{12}+Z_{23}}{Z_{12}(Z_{13}+Z_{23})} \Big| Z_{13} < Z_{12}\right) \cdot
  \Pr (Z_{13} < Z_{12})}_{b}.
\end{eqnarray*}
A simple calculation shows that $a=\frac{1}{2} \left( 1 +
  e^{-\frac{2}{S}} \right) - e^{-\frac{1}{S}}$. Conditioned on the
event $\{ Z_{13} < Z_{12}\}$, $\left\{ S \leq \frac{1}{Z_{12}} \right\}
  \cup \left\{ S \leq \frac{1}{Z_{13}+Z_{23}}\right\} \subseteq
  \left\{ S \leq \frac{Z_{12}+Z_{23}}{Z_{12}(Z_{13}+Z_{23})}\right\}$. Hence
\begin{eqnarray*}
b & \geq &
\Pr \left( \left\{ S \leq \frac{1}{Z_{12}} \right\}
  \cup \left\{ S \leq \frac{1}{Z_{13}+Z_{23}}\right\}
  \Big| Z_{13} < Z_{12}\right) \cdot \Pr (Z_{13} < Z_{12}) \\
&=&
 \Pr (Z_{13} < Z_{12}) -
  \Pr \left( \left\{ S > \frac{1}{Z_{12}} \right\}
  \cap \left\{ S > \frac{1}{Z_{13}+Z_{23}}\right\}
  \cap \{Z_{13} < Z_{12} \} \right) \\
&=&
  \int_{0}^{\infty} \left[ \Pr(Z_{12}>z) - \Pr\left( Z_{12} >
        \max\left\{z,\frac{1}{S}\right\}\right) \cdot \Pr\left( Z_{23} >
          \frac{1}{S} - z\right) \right] e^{-z} dz \\
&=&
  \int_{0}^{\frac{1}{S}} \left( e^{-z} - e^{-\frac{1}{S}} \cdot
    e^{-\frac{1}{S}+z}\right) e^{-z} dz +
  \cdot \int_{\frac{1}{S}}^{\infty} \left( e^{-z} - e^{-z} \cdot 1
    \right) e^{-z} dz \\
&=&
  \frac{1}{2}\left(1-e^{-\frac{2}{S}}\right) - \frac{1}{S} e^{-\frac{2}{S}}.
\end{eqnarray*}
It is also easy to see that $\lim_{S\rightarrow \infty}
\frac{a+b}{1/S^2} = 1.5$.

\item By Corollary~\ref{thm:b1asymp}, $B_1(K) \leq
  \lim_{K'\rightarrow \infty} B_1(K')$ for all $K>0$. Note that when
  $K$ is sufficiently large, $B_1(K) = \frac{\tilde S_1(K)}{e^K-1}$.
  Now instead of choosing the optimal $t^*$ in (\ref{e:tS1}), we
  choose $t_1=1/2$ and normalize the suboptimal solution by the factor
  $e^K-1$. Taking limit as $K\rightarrow \infty$, we get
  \[
  \tilde B_1 = \left\{
  \begin{array}{ll}
    \displaystyle
    \sqrt{\frac{1}{Z_{23}} \left(\frac{1}{Z_{13}} -\frac{1}{Z_{12}} \right)}
    +
    \sqrt{\frac{1}{Z_{12}} \left(\frac{1}{Z_{13}} -\frac{1}{Z_{23}} \right)}
      &
    \mbox{~if~} Z_{13} < M_H(Z_{12},Z_{23})\\
    \displaystyle
    \frac{1}{Z_{13}} &
    \mbox{~if~} Z_{13} \geq M_H(Z_{12},Z_{23}).
  \end{array} \right.
  \]
  Obviously, $\lim_{K'\rightarrow \infty} B_1(K') \leq \tilde B_1$
  because of the suboptimality of the choice $t_1=1/2$.  Thus,
  $P_1(K,S) \leq \Pr (S\leq \tilde B_1)$. Moreover, when $Z_{13} <
  M_H(Z_{12},Z_{23})$,
  \begin{eqnarray*}
  \sqrt{\frac{1}{Z_{23}} \left(\frac{1}{Z_{13}} -\frac{1}{Z_{12}} \right)}
  +
  \sqrt{\frac{1}{Z_{12}} \left(\frac{1}{Z_{13}} -\frac{1}{Z_{23}} \right)}
  &=&
  \frac{2 \left( \frac{1}{2} \sqrt{\frac{Z_{12}}{Z_{13}} -1} +
  \frac{1}{2} \sqrt{\frac{Z_{23}}{Z_{13}} -1} \right) }{\sqrt{Z_{12}Z_{23}}} \\
  & \leq &
  \frac{2\sqrt{\frac{Z_{12}+Z_{23}}{2Z_{13}} - 1}}{\sqrt{Z_{12}Z_{23}}} \\
  & < & 
  \sqrt{\frac{2}{Z_{13}}} \cdot \sqrt{\frac{1}{Z_{12}} + \frac{1}{Z_{23}}},
  \end{eqnarray*}
  where the second line is due to the concavity of the square-root
  function. Hence
  \begin{eqnarray*}
  P_1(K,S) &\leq &
  \underbrace{\Pr\left( S\leq \frac{1}{Z_{13}} \Big| \frac{1}{Z_{12}}
      + \frac{1}{Z_{23}} \geq \frac{1}{Z_{13}} \right) \cdot
      \Pr\left(\frac{1}{Z_{12}} + \frac{1}{Z_{23}} \geq \frac{1}{Z_{13}}
      \right)}_{a} \\
  & & +
  \underbrace{\Pr\left( S <
  \sqrt{\frac{2}{Z_{13}}} \cdot \sqrt{\frac{1}{Z_{12}} + \frac{1}{Z_{23}}}
    \right) \cdot
      \Pr\left(\frac{1}{Z_{12}} + \frac{1}{Z_{23}} < \frac{1}{Z_{13}}
      \right)}_{b}.
  \end{eqnarray*}
  
  By Claim~\ref{c1},
  \begin{eqnarray*}
  a &=& \int_{0}^{\frac{1}{S}} \Pr\left( \frac{1}{Z_{12}}
      + \frac{1}{Z_{23}} \geq \frac{1}{z} \right) e^{-z} dz \\
    &=&
      \int_{0}^{\frac{1}{S}} \left[ 1- 2z K_1(2z) e^{-2z} \right] e^{-z} dz \\
    &=&
      1- e^{\frac{1}{S}} - \int_{0}^{\frac{1}{S}} 2z K_1(2z) e^{-3z} dz,
  \end{eqnarray*}
  and
  \begin{eqnarray*}
  b &=& \int_{0}^{\frac{\sqrt{2}}{S}} \Pr\left( \frac{S^2z}{2} < 
    \frac{1}{Z_{12}}
      + \frac{1}{Z_{23}} < \frac{1}{z} \right) e^{-z} dz \\
    &=&
    \int_{0}^{\frac{\sqrt{2}}{S}} \left[ 2zK_1(2z)e^{-2z} - \frac{4}{S^2 z} 
   K_1\left(\frac{4}{S^2 z}\right) e^{-\frac{4}{S^2 z}} \right] e^{-z} dz.
  \end{eqnarray*}

  Now by repeated uses of L'Hospital's rule, we have
  \begin{eqnarray*}
  \lim_{S\rightarrow \infty} \frac{a}{1/S^2} & = &
  \lim_{u\rightarrow 0}
  \frac{1- e^{u} - \int_{0}^{u} 2z K_1(2z) e^{-3z} dz}{u^2} \\
  & = &
  \lim_{u\rightarrow 0} e^{-u} \cdot
  \frac{1- 2u K_1(2u) e^{-2u}}{2u} \\
  &=&
  \lim_{u\rightarrow 0} 2u [ K_1(2u) + K_0(2u)] e^{-2u} \\
  &=& 1,
  \end{eqnarray*}
  where the third equality is due to the fact that the derivative of
  $-uK_1(u)e^{-u}$ is $u[K_1(u)+K_0(u)]e^{-u}$ \cite{Hasna03}, and the
  last equality is due to the facts that $\lim_{u\rightarrow 0}
  uK_1(u) = 1$ and $\lim_{u\rightarrow 0} uK_0(u) = 0$
  \cite{Tranter68}. To find the asymptotic order of $b$, let us write
  \[
  c = 
    \int_{0}^{\frac{\sqrt{2}}{S}} \left[ 2zK_1(2z)e^{-2z} - \frac{4}{S^2 z} 
   K_1\left(\frac{4}{S^2 z}\right) e^{-\frac{4}{S^2 z}} \right] dz.
  \]
  First, we note that
  \[
    e^{-\frac{\sqrt{2}}{S}} c  \leq b \leq c. 
  \]
  Then again by repeated applications of L'Hospital's rule, we have
  \begin{eqnarray*}
  \lim_{S\rightarrow \infty} \frac{c}{\log(S)/S^2} & = &
  \lim_{u\rightarrow 0} \frac{
    \int_{0}^{\sqrt{2u}} \left[ 2zK_1(2z)e^{-2z} - \frac{4u}{z} 
    K_1\left(\frac{4u}{z}\right) e^{-\frac{4u}{z}} \right] dz}{
    -\frac{1}{2} u \log u }\\
  &=&
   8 \lim_{u\rightarrow 0} \frac{
    \int_{2\sqrt{2u}}^{\infty} [ K_1(x) +K_0(x)] e^{-x} dx}{
    -\log u - 1 }\\
  &=&
   4 \lim_{u\rightarrow 0} 2\sqrt{2u}
    \left[ K_1(2\sqrt{2u}) +K_0(2\sqrt{2u}) \right] e^{-2\sqrt{2u}} \\
  &=& 4,
  \end{eqnarray*}
  where the second equality is obtained by a change of integration
  variable after the use of L'Hospital's rule. As a consequence, $
  \lim_{S\rightarrow \infty} \frac{b}{\log(S)/S^2} = 4$.
  
\item By Corollary~\ref{thm:b1asymp}, $B_1(K) \geq \lim_{K'\rightarrow
    0} B_1(K') = \min\left\{\displaystyle \frac{1}{Z_{13}},
    \frac{1}{Z_{23}}+\frac{1}{Z_{12}} \right\}$ for all $K>0$. In
  addition, the bound is achieved as $K$ approaches zero. Thus
  \begin{eqnarray*}
  P_1(K,S) &\geq &
  \Pr\left( S\leq\min\left\{\displaystyle \frac{1}{Z_{13}},
    \frac{1}{Z_{23}}+\frac{1}{Z_{12}} \right\} \right) \\
  &=& 
  \Pr\left( S \leq \frac{1}{Z_{13}} \right) \cdot 
  \Pr\left( S \leq \frac{1}{Z_{23}} + \frac{1}{Z_{12}} \right) \\
  &=&
  \left[ 1 - e^{-\frac{1}{S}} \right]
  \left[1 - \frac{2}{S} K_1\left(\frac{2}{S}\right) e^{-\frac{2}{S}} \right],
  \end{eqnarray*}
  where the last line is due to Claim~\ref{c1} and the bound is
  achieved as $K\rightarrow 0$ by monotone convergence. Moreover,
  \begin{eqnarray*}
  \lim_{S\rightarrow \infty} \frac{
  \left[1 - \frac{2}{S} K_1\left(\frac{2}{S}\right) e^{-\frac{2}{S}} \right]
  \left[ 1 - e^{-\frac{1}{S}} \right]}{1/S^2} & = &
  \lim_{u\rightarrow 0} \frac{1-2uK_1(2u)e^{-2u}}{u} \cdot
  \lim_{u\rightarrow 0} \frac{1-e^{-u}}{u} \\
  &=& 2 \cdot 1.
  \end{eqnarray*}

\item By Corollary~\ref{thm:b2asymp} and similar to part 3), we have
  \begin{eqnarray*}
  P_2(K,S) &\leq &
  \underbrace{\Pr\left( S\leq \frac{1}{Z_{13}} \Big| \frac{1}{Z_{12}}
      + \frac{1}{Z_{13}+Z_{23}} \geq \frac{1}{Z_{13}} \right) \cdot
      \Pr\left(\frac{1}{Z_{12}} + \frac{1}{Z_{13}+Z_{23}} \geq \frac{1}{Z_{13}}
      \right)}_{a} \\
  & & +
  \underbrace{\Pr\left( S < \sqrt{\frac{2}{Z_{13}}} \cdot
      \sqrt{\frac{1}{Z_{12}} + \frac{1}{Z_{13}+Z_{23}}} \right) \cdot
      \Pr\left(\frac{1}{Z_{12}} + \frac{1}{Z_{13}+Z_{23}} < \frac{1}{Z_{13}}
      \right)}_{b}.
  \end{eqnarray*}
  By Claim~\ref{c1},
  \begin{eqnarray*}
  a &=& \int_{0}^{\frac{1}{S}} \Pr\left( \frac{1}{Z_{12}}
       + \frac{1}{Z_{23}+z} \geq \frac{1}{z} \right) e^{-z} dz \\
    &=&
      1- e^{\frac{1}{S}} - \int_{0}^{\frac{1}{S}} 2z K_1(2z) e^{-2z} dz,
  \end{eqnarray*}
  and
  \begin{eqnarray*}
  b &=& \int_{0}^{\frac{\sqrt{2}}{S}} \Pr\left( \frac{S^2z}{2} < 
    \frac{1}{Z_{12}}
      + \frac{1}{Z_{23}+z} < \frac{1}{z} \right) e^{-z} dz \\
    &=&
    \int_{0}^{\frac{\sqrt{2}}{S}} \left[ 2zK_1(2z)e^{-2z} - \frac{4}{S^2 z} 
   K_1\left(\frac{4}{S^2 z}\right) e^{-\frac{4}{S^2 z}} \right] dz.
  \end{eqnarray*}

  Now by repeated uses of L'Hospital's rule, we have
  \begin{eqnarray*}
  \lim_{S\rightarrow \infty} \frac{a}{1/S^2} & = &
  \lim_{u\rightarrow 0}
  \frac{1- e^{u} - \int_{0}^{u} 2z K_1(2z) e^{-2z} dz}{u^2} \\
  & = &
  \lim_{u\rightarrow 0} \frac{e^{-u}- 2u K_1(2u) e^{-2u}}{2u} \\
  &=& 1/2.
  \end{eqnarray*}
  As derived in part 3), $\displaystyle \lim_{S\rightarrow \infty}
  \frac{b}{\log(S)/S^2} =4$.

\item By Corollary~\ref{thm:b2asymp} and similar to part 4), we have
  \begin{eqnarray*}
  P_2(K,S) &\geq &
  \Pr\left( S\leq\min\left\{\displaystyle \frac{1}{Z_{13}},
    \frac{1}{Z_{13}+Z_{23}}+\frac{1}{Z_{12}} \right\} \right) \\
  &=& 
  \int_{0}^{\frac{1}{S}} \Pr\left(
    S \leq \frac{1}{Z_{12}} + \frac{1}{Z_{23}+z} \right) e^{-z} dz\\
  &=&
  \int_{0}^{\frac{1}{S}} \left[ e^{-z}
   - \frac{2}{S} K_1\left(\frac{2}{S}\right) e^{-\frac{2}{S}} \right] dz\\
  &=&
  1 - e^{-\frac{1}{S}}  
   - \frac{2}{S^2} K_1\left(\frac{2}{S}\right) e^{-\frac{2}{S}},
  \end{eqnarray*}
  where the equality in the third line is established by
  Claim~\ref{c1} and the bound is achieved as $K\rightarrow 0$ by
  monotone convergence.
  Moreover,
  \begin{eqnarray*}
  \lim_{S\rightarrow \infty} \frac{1 - e^{-\frac{1}{S}} -
    \frac{2}{S^2} K_1\left(\frac{2}{S}\right) e^{-\frac{2}{S}}}{1/S^2}
  & = &
  \lim_{u\rightarrow 0} \frac{1-e^{-u}-2u^2K_1(2u)e^{-2u}}{u^2} \\
  & = &
  \lim_{u\rightarrow 0} \frac{e^{-u} - 2uK_1(2u)e^{-2u}}{2u} +
  \lim_{u\rightarrow 0} 2u [K_1(2u)+K_0(2u)]e^{-2u} \\
  & = &
  \lim_{u\rightarrow 0} \frac{-e^{-u} - 4u[K_1(2u)+K_0(2u)]e^{-2u}}{2} + 1\\
  &=&  1/2 + 1 = 3/2.
  \end{eqnarray*}
\end{enumerate}

\subsection{Proof of Theorem~\ref{thm:ec}}
\label{app:ec}
To prove part 1) of the theorem, we employ the Fano inequality as in
\cite{Zheng03}. For the remaining parts, the achievability proofs are
based on extending the Feinstein lemma
\cite{Feinstein54,Han03,Verdu94} to the various cases of interest.

\begin{enumerate}
\item Suppose that $K$ is $(\varepsilon,P_t)$-achievable. Hence, for
  any $0 < \gamma < K$, there is a sequence of
  $(n,M_n,\varepsilon_n,P_n)$-codes satisfying $\varepsilon_n \leq
  \varepsilon + \gamma$, $ \frac{1}{n}\log M_n \geq K - \gamma$, and
  $P_n \leq P_t + \gamma$ a.s. for all sufficiently large $n$.  Let
  $M$ be the uniform random variable representing the message being
  set from the source to destination. Since $M$ is independent of $Z$,
  conditioned on the link realization $Z=(Z_{13},Z_{12},Z_{23})$, by
  the Fano inequality,
  \begin{eqnarray}
  \Pr (\mathrm{error} | Z=(Z_{13},Z_{12},Z_{23}))
  & \geq &
  1 - \frac{1}{\log M_n} [I(M;Y^n | Z=(Z_{13},Z_{12},Z_{23}))
  + 1] \nonumber \\
  & \geq &
  1 - \frac{1}{n(K-\gamma)} [I(M;Y^n| Z=(Z_{13},Z_{12},Z_{23}))
  + 1],
  \label{e:fano}
  \end{eqnarray}
  for all sufficiently large $n$.  Since the relay channel is
  memoryless conditioned on $Z=(Z_{13},Z_{12},Z_{23})$, by the same
  argument as in the proof of Theorem~4 in \cite{Cover79}
  \begin{eqnarray}
  \lefteqn{I(M;Y^n|Z=(Z_{13},Z_{12},Z_{23}))} \nonumber \\
  &\leq& n \cdot
  \min\{I(X_1,X_2;Y |Z=(Z_{13},Z_{12},Z_{23})),
  I(X_1;Y,Y_1|Z=(Z_{13},Z_{12},Z_{23}))\}
  \nonumber  \\
  & \leq & n \cdot R_{\mathrm{lb}} (Z_{13},Z_{12},Z_{23};P_t+\gamma)
  \label{e:cIb}
  \end{eqnarray}
where
\begin{eqnarray*}
  \lefteqn{R_{\mathrm{lb}} (Z_{13},Z_{12},Z_{23};P_t+\gamma)} \\
  &=& 
  \max_{0\leq \alpha,\beta \leq 1} \min\left\{
  C\left( \frac{P_t + \gamma}{N_0 W}
    \left[ \alpha Z_{13} + (1-\alpha) Z_{23} + 2 
      \sqrt{(1-\beta)\alpha (1-\alpha) Z_{13}Z_{23}}\right]\right), \right.
  \\
  && ~~~~~~~~~ 
  C\left.\left( \frac{P_t+\gamma}{N_0 W} \alpha \beta [Z_{12}+Z_{13}]
    \right) \right\}.
  \end{eqnarray*}
  The two conditional mutual information terms on the right hand side
  of the first inequality in (\ref{e:cIb}) are based on the
  element-wise conditional pdf $p_{Y, Y_1| X_1, X_2,
    Z}(y,y_1|x_1,x_2)$ and conditional input pdf
  $p_{X_1,X_2|Z}(x_1,x_2)$ based on the code as in \cite{Cover79} and
    the second inequality is due to the fact these mutual information
    terms are maximized by independent Gaussian inputs $X_1$ and $X_2$
    \cite{Host05} and $P_n \leq P_t + \gamma$ for sufficiently large
    $n$.
  
  Now putting (\ref{e:cIb}) into (\ref{e:fano}) and noting that
  $\Pr(\mathrm{error}|Z) \geq 0$, we have
  \[
  \varepsilon + \gamma \geq \varepsilon_n = E[\Pr(\mathrm{error}|Z)]
  \geq E\left[\max\left\{1 - \frac{
        R_{\mathrm{lb}}(Z_{13},Z_{12},Z_{23};P_t+\gamma)}{K-\gamma}
      - \frac{1}{n(K-\gamma)},0\right\}\right]
  \]
  for sufficiently all large $n$.  Since $\gamma$ is arbitrary,
  \[
  \varepsilon \geq E\left[ \max\left\{1 -
      \frac{R_{\mathrm{lb}}(Z_{13},Z_{12},Z_{23};P_t)}{K},0\right\}
  \right] \geq \delta \Pr\left(
    R_{\mathrm{lb}}(Z_{13},Z_{12},Z_{23};P_t) < (1-\delta)K \right),
  \]
  for all $0 < \delta < 1$.  From Theorem~\ref{thm:fd}, we have then $
  \varepsilon \geq \delta P_{\mathrm{lb}} \left(K,
    \frac{e^K-1}{e^{(1-\delta) K} - 1} S \right)$.

\item We employ the approach of block encoding and parallel Gaussian
  channel decoding suggested in \cite{Xie04}. First we divide a time
  slot into $\tilde n = \sqrt{n}$ sub-slots\footnote{For notational
    simplicity, we assume $\sqrt{n}$ is an integer with no loss of
    generality.}.  We are to send $\tilde n - 1$ messages, each coming
  from one of $M_{\tilde n}$ possibilities, in the whole time slot.
  Thus in each sub-slot, we have $\tilde n=\sqrt{2W}$ symbols.  Let
  $\mathcal{N}(\mu,\sigma^2)$ be the Gaussian distribution with mean
  $\mu$ and variance $\sigma^2$.  Consider the following code
  construction.

  \setcounter{paragraph}{0}
  \paragraph{Codebook generation}
  Independently generate $M_{\tilde n}$ $\tilde n$-element vectors
  $u_1, u_2, \ldots, u_{M_{\tilde n}}$ with all elements in the
  vectors distributed according to i.i.d. $\mathcal{N}(0,1)$
  Similarly, independently generate $M_{\tilde n}$ $\tilde n$-element
  vectors $v_1, v_2, \ldots, v_{M_{\tilde n}}$ with all elements in
  the vectors distributed according to i.i.d.  $\mathcal{N}(0,1)$.
  
  \paragraph{Encoding}
  Let the $k$th message be $w_k$, for $k=1,2,\ldots,\tilde n -1$, and
  $w_0=1$, which is known to the relay and destination beforehand. Let
  $0 \leq \beta \leq 1$. In the $k$th sub-slot, the source sends
  $\sqrt{\beta P_1} v_{w_k} + \sqrt{(1-\beta)P_1} u_{w_{k-1}}$ and the
  relay sends $\sqrt{P_2} u_{\hat w_{k-1}}$, where $\hat w_{k-1}$ is
  the estimate of $w_{k-1}$ that the relay obtains based on the signal
  that it receives in the $(k-1)$th sub-slot. In above, $P_1 = \alpha
  P_t/2W$ and $P_2 = (1-\alpha)P_t/2W$, where $0\leq \alpha \leq 1$..
  
  When the code defined above is employed, we can rewrite the
  relationship between the output and input symbols described by
  (\ref{e:model}) as
  \begin{eqnarray}
  Y^{\tilde n,k}&=& \sqrt{\beta Z_{13} P_1} \tilde X_1^{\tilde n,k} + 
  \left[ \sqrt{(1-\beta)Z_{13}P_1} + \sqrt{Z_{23}P_2} \right]
  \tilde X_2^{\tilde n,k} + N^{\tilde n,k}
  \nonumber \\
  Y_1^{\tilde n,k} &=& \sqrt{\beta Z_{12}P_1} \tilde X_1^{\tilde n,k} +
  \sqrt{(1-\beta)Z_{12} P_1} \tilde X_2^{\tilde n,k} + N_1^{\tilde n,k},
  \label{e:model2}
  \end{eqnarray}
  where the superscript $k$ denotes the $k$th sub-slot.  Hence the
  relay channel can be alternatively specified by the conditional pdf
  $p_{Y^{\tilde n,k}, Y_1^{\tilde n,k}| \tilde X_1^{\tilde n,k},
    \tilde X_2^{\tilde n,k}, \alpha, \beta, Z}(y^{\tilde
    n,k},y_1^{\tilde n,k}|\tilde x_1^{n,k},\tilde x_2^{\tilde n,k})$
  with $(\tilde X_1^{n,k}, \tilde X_2^{n,k})$. Note that since the
  channel state information is available at the source and relay,
  $\alpha$ and $\beta$ are functions of $Z$ in general. This
  corresponds to the application of flow control.

  \paragraph{Decoding}
  Consider decoding of the message $w_{k}$ ($k=1,2,\ldots,\tilde n-1$)
  under the assumption that the previous message $w_{k-1}$ has been
  correctly decoded, and hence is known, at both the relay and
  destination. Fix any $\gamma >0$, define the sets
  \begin{eqnarray*}
  T_1^{\tilde n,k}(\alpha,\beta,Z) &=&
  \Bigg\{ \left(\tilde x_1^{n,k-1},\tilde x_2^{\tilde n,k-1},
      \tilde x_1^{\tilde n,k},\tilde x_2^{\tilde n,k},y^{\tilde n,k-1},
      y^{\tilde n,k}\right) : \\
    & & ~~~~
    \frac{1}{\tilde n} \log \frac{p_{Y^{\tilde n,k-1},Y^{\tilde n,k}|
        \tilde X_1^{\tilde n,k-1}, \tilde X_2^{\tilde n,k-1},
        \tilde X_2^{\tilde n,k}, \alpha, \beta, Z}
      (y^{\tilde n,k-1},y^{\tilde n,k}|\tilde x_1^{\tilde n,k-1},
      \tilde x_2^{\tilde n,k-1}, \tilde x_2^{\tilde n,k})}%
    {p_{Y^{\tilde n,k-1},Y^{\tilde n,k}| \tilde X_2^{\tilde n,k-1}, \alpha,
        \beta, Z} 
      (y^{\tilde n,k-1},y^{\tilde n,k}|\tilde x_2^{\tilde n,k-1})} \\
    & & ~~~~ > \frac{1}{\tilde n} \log M_{\tilde n} + \gamma \Bigg\} \\
  T_2^{n,k}(\alpha, \beta, Z) &=&
  \left\{ \left(\tilde x_1^{\tilde n,k},\tilde x_2^{\tilde n,k},
      y_1^{\tilde n,k}\right) :
    \frac{1}{\tilde n} \log \frac{p_{Y_1^{\tilde n,k}|
        \tilde X_1^{\tilde n,k}, \tilde X_2^{\tilde n,k}, \alpha, \beta, Z}
      (y_1^{\tilde n,k}|\tilde x_1^{\tilde n,k},\tilde x_2^{\tilde n,k})}%
    {p_{Y_1^{\tilde n,k}| \tilde X_2^{\tilde n,k}, \alpha, \beta, Z} 
      (y_1^{\tilde n,k}|\tilde x_2^{\tilde n,k})} >
    \frac{1}{\tilde n} \log M_{\tilde n} + \gamma \right\} 
  \end{eqnarray*}
  In the $k$th sub-slot, the relay outputs $\hat w_{k} = i$ if and
  only if there is a unique $i$ (from $1$ to $M_{\tilde n}$) such that
  $(v_i,u_{w_{k-1}},y_1^{\tilde n,k}) \in T_2^{\tilde
    n,k}(\alpha,\beta,Z)$. This allows the encoding steps mentioned
  above to continue in the $(k+1)$th sub-slot. In the $(k+1)$th
  sub-slot, the destination outputs $\hat w_{k} = i$ if and only if
  there a unique $i$ such that $(v_i,u_{w_{k-1}},u_i,y^{\tilde
    n,k},y^{\tilde n,k+1}) \in T_1^{\tilde n,k+1}(\alpha,\beta,Z)$.

  \paragraph{Error analysis}
  Let $\varepsilon_{n}$ be the average\footnote{The error probability
    defined here is averaged over all Gaussian codes constructed as
    described. Thus there is a Gaussian code that gives at least the
    same error performance.}  error probability of decoding the whole
  time slot and $F_k$ be the event of erroneous decoding in the $k$th
  sub-slot, for $k=1,2,\ldots,\tilde n$. Then
  \[
  \varepsilon_{n} = \Pr\left( \cup_{k=1}^{\tilde n} F_k \right) =
  \Pr\left(\cup_{k=1}^{\tilde n} \left\{ F_k \cap
      \left[\cap_{l=1}^{k-1} F_{l}^c \right]\right\}\right).
  \]
  
  Consider the event $F_k \cap \left( \cap_{l=1}^{k-1} F_{l}^c
  \right)$. The message $w_{k-2}$ is correctly decoded, and hence is
  known, at the relay and destination, while the message $w_{k-1}$ is
  corrected decoded, and hence is known, at the relay.  By symmetry of
  the code generated, we can assume $w_{k-2}=w_{k-1} = w_{k} = 1$ with
  no loss of generality. For $i=1,2,\ldots,M_{\tilde n}$, write
  $E_{i,k}^{1} = \left\{(v_i,u_1,u_i,y^{\tilde n,k-1},y^{\tilde n,k})
    \in T_1^{\tilde n,k}(\alpha,\beta,Z) \right\}$ and $E_{i,k}^2 =
  \left\{(v_i,u_1,y_1^{\tilde n,k}) \in T_2^{\tilde n,k}
    (\alpha,\beta,Z) \right\}$.  Then
  \[
  F_k \cap \left( \cap_{l=1}^{k-1} F_{l}^c \right) = (E_{1,k}^1)^c
  \cup (E_{1,k}^2)^c \cup \left( \cup_{i=2}^{M_{\tilde n}} E_{i,k}^1
  \right) \cup \left( \cup_{i=2}^{M_{\tilde n}} E_{i,k}^2 \right).
  \]
  
  Now for $i=2,3,\ldots,M_{\tilde n}$,
  \begin{eqnarray*}
    \Pr (E_{i,k}^1) &=&
    \int_{T_1^{\tilde n,k}(\alpha,\beta,Z)} 
    p_{\tilde X_1^{\tilde n,k-1},\tilde X_2^{\tilde n,k-1},
      \tilde X_2^{\tilde n,k},Y^{\tilde n,k-1},Y^{\tilde n,k}|
      \alpha,\beta,Z}
    (v_i,u_1,u_i,y^{\tilde n,k-1},y^{\tilde n,k}) \\
    &=&
    \int_{T_1^{\tilde n,k}(\alpha,\beta,Z)}
    p_{Y^{\tilde n,k-1},Y^{\tilde n,k}|\tilde X_2^{\tilde n,k-1},
      \alpha,\beta,Z}(y^{\tilde n,k-1},y^{\tilde n,k}|u_1) \\
    & & ~~~~~~~~~~~~~~ \cdot 
    p_{\tilde X_1^{\tilde n,k-1},\tilde X_2^{\tilde n,k}|
      \tilde X_2^{\tilde n,k-1},\alpha,\beta,Z} (v_i,u_i|u_1)
    p_{\tilde X_2^{\tilde n,k-1}|\alpha,\beta,Z} (u_1) \\
    & \leq &
    \frac{e^{-\tilde n\gamma}}{M_{\tilde n}} 
    \int_{T_1^{\tilde n,k}(\alpha,\beta,Z)}
    p_{Y^{\tilde n,k-1},Y^{\tilde n,k}|\tilde X_1^{\tilde n,k-1},
      \tilde X_2^{\tilde n,k-1},\tilde X_2^{\tilde n,k},\alpha,\beta,Z}
    (y^{\tilde n,k-1},y^{\tilde n,k}|v_i,u_1,u_i) \\
    & & ~~~~~~~~~~~~~~~~~~~~~ \cdot
    p_{\tilde X_1^{\tilde n,k-1},\tilde X_2^{\tilde n,k-1},
      \tilde X_2^{\tilde n,k}|\alpha,\beta,Z} (v_i,u_1,u_i) \\
    & \leq &
    \frac{e^{-\tilde n\gamma}}{M_{\tilde n}} 
  \end{eqnarray*}
  where the independence between $Y^{\tilde n,k}$ and $u_i$ in the
  second line is due to the fact that $Y^{\tilde n,k}$ and $u_i$ are
  jointly Gaussian and uncorrelated, and the inequality on the third
  line follows from the definition of $T_1^{\tilde
    n,k}(\alpha,\beta,Z)$.  Similarly, we can employ the definition of
  $T_2^{\tilde n,k}(\alpha,\beta,Z)$ to show that $\Pr (E_{i,k}^2)
  \leq \frac{e^{-\tilde n\gamma}}{M_{\tilde n}}$ for $i = 2, 3,
  \ldots, M_{\tilde n}$.  Hence
  \begin{eqnarray*}
  \varepsilon_n &\leq& \Pr \left( \cup_{k=1}^{\tilde n}
    \left[(E_{1,k}^1)^c \cup (E_{1,k}^2)^c \right] \right) +
  \sum_{k=1}^{\tilde n} \sum_{i=2}^{M_{\tilde n}} \Pr (E_{i,k}^1) +
  \sum_{k=1}^{\tilde n} \sum_{i=2}^{M_{\tilde n}} \Pr(E_{i,k}^2) \\
  &\leq&  
  \Pr \left( \cup_{k=1}^{\tilde n}
    \left[(E_{1,k}^1)^c \cup (E_{1,k}^2)^c \right] \right) +
    2\tilde n e^{-\tilde n\gamma}.
  \end{eqnarray*}
  
  Putting the codes in the $\tilde n$ sub-slots together, we obtain a
  sequence of $(n,M_{\tilde n}^{\tilde n-1},\varepsilon_{n},
  P_n)$-codes over the time slot with $\varepsilon_{n}$ and $P_n$
  respectively satisfying $\limsup_{n\rightarrow\infty}
  \varepsilon_{n} \leq \limsup_{n\rightarrow\infty} \Pr \left(
    \cup_{k=1}^{\tilde n} \left[(E_{1,k}^1)^c \cup (E_{1,k}^2)^c
    \right] \right)$ and $\limsup_{n\rightarrow\infty} P_{n} = P_t$
  a.s. 
  
  Further note that as $n$ (and hence $\tilde n$) becomes large
  \begin{eqnarray*}
  \lefteqn{\frac{1}{\tilde n} \log 
    \frac{p_{Y^{\tilde n,k-1},Y^{\tilde n,k}|\tilde X_1^{n,k-1},
     \tilde X_2^{\tilde n,k-1}, \tilde X_2^{\tilde n,k}, \alpha,\beta,Z}
      (y^{\tilde n,k-1},y^{\tilde n,k}|\tilde x_1^{\tilde n,k-1},
      \tilde x_2^{\tilde n,k-1},\tilde x_2^{\tilde n,k})}%
    {p_{Y^{\tilde n,k-1},Y^{\tilde n,k}|
        \tilde X_2^{\tilde n,k-1},\alpha,\beta, Z} 
      (y^{\tilde n,k-1},y^{\tilde n,k}|\tilde x_2^{\tilde n,k-1})}} \\
    &~~~~~~~~~~~~~~~~\stackrel{\mathrm{a.s.}}{\longrightarrow} &  
  C\left( \frac{P_t}{N_0 W} \left[ \alpha Z_{13} + (1-\alpha) Z_{23} + 2 
      \sqrt{(1-\beta)\alpha (1-\alpha) Z_{13}Z_{23}}\right]\right) \\
  \lefteqn{\frac{1}{\tilde n} \log \frac{p_{Y_1^{\tilde n,k}|
        \tilde X_1^{\tilde n,k}, \tilde X_2^{\tilde n,k},\alpha,\beta, Z}
      (y_1^{\tilde n,k}|\tilde x_1^{\tilde n,k},\tilde x_2^{\tilde n,k})}%
    {p_{Y_1^{\tilde n,k}| \tilde X_2^{\tilde n,k},\alpha,\beta, Z} 
      (y_1^{n,k}|x_2^{n,k})}  
    \stackrel{\mathrm{a.s.}}{\longrightarrow} 
    C\left(\frac{\alpha\beta Z_{13} P_t}{N_0W} \right)}
  \end{eqnarray*}
  for all $k=1,2,\ldots,\tilde n$, when the inputs symbols are
  Gaussian distributed as described in the code generation step above.
  
  Now set $M_{\tilde n} = e^{\frac{\tilde n^2}{\tilde n -1}K}$ and
  choose $\alpha$ and $\beta$ so that the maximum rate
  $R_{\mathrm{DF}}$ in Appendix~\ref{app:fd} is achieved. Since the
  choice of $\gamma > 0$ is arbitrary, the code construction argument
  above shows the existence of a sequence of
  $(n,M_{n},\varepsilon_{n}, P_{n})$-codes over the time slot
  satisfying $\liminf_{n\rightarrow\infty} \frac{1}{n} \log M_{n} =
  K$, $\limsup_{n\rightarrow\infty} \varepsilon_{n} \leq
  \limsup_{\tilde n\rightarrow\infty} \Pr \left(R_{\mathrm{DF}} \leq
    \frac{\tilde n}{\tilde n-1}K\right) = \Pr \left(R_{\mathrm{DF}}
    \leq K \right)$, and $\limsup_{n\rightarrow\infty} P_{n} = P_t$
  a.s.~. From Theorem~\ref{thm:fd}, $\Pr\left(R_{\mathrm{DF}} \leq K
  \right) = P_\mathrm{DF}(K,S)$.  Therefore if $ P_\mathrm{DF} (K,S)
  \leq \varepsilon$, then the rate $K$ is
  $(\varepsilon,P_t)$-achievable.
  
\item We construct a code that conforms to HDP1. Fix $0<t_1<1$ and
  $t_2=1-t_1$. Write $n_1=\lfloor t_1 n \rfloor$ and $n_2 =n - n_1$.

  \setcounter{paragraph}{0}
  \paragraph{Codebook generation}
  Independently generate $M_{n}^1$ $n_1$-element vectors $u_1, u_2,
  \ldots, u_{M_n^1}$ with all elements in the vectors distributed
  according to i.i.d. $\mathcal{N}(0,1)$.  Independently generate
  $M_{n}^2$ $n_1$-element vectors $s_1^1, s^1_2, \ldots, s^1_{M_n^2}$
  with all elements in the vectors distributed according to i.i.d.
  $\mathcal{N}(0,1)$.  Similarly, independently generate $M_n^3$
  $n_2$-element vectors $v_1, v_2, \ldots, v_{M_n^3}$ with all
  elements in the vectors distributed according to i.i.d.
  $\mathcal{N}(0,1)$.  Independently generate $M_{n}^2$ $n_2$-element
  vectors $s_1^2, s^2_2, \ldots, s^2_{M_n^2}$ with all elements in the
  vectors distributed according to i.i.d.  $\mathcal{N}(0,1)$.
  
  \paragraph{Encoding}
  Let $M_n = M_n^1 \times M_n^2 \times M_n^3$. A message $w$, with
  value from $1$ to $M_n$, can be indexed by the triple $(i,j,k)$,
  where $i$ ranges from $1$ to $M_n^1$, $j$ ranges from $1$ to
  $M_n^2$, and $k$ ranges from $1$ to $M_n^3$. Also we divide a time
  slot with $n$ symbols into two sub-slots: the first with $n_1$
  symbols and the second with $n_2$ symbols. In the first sub-slot,
  the source sends $\sqrt{\frac{\alpha P_t}{2W}} u_i +
  \sqrt{\frac{(1-\alpha)P_t}{2W}} s_j^1$, where $0 \leq \alpha \leq
  1$.  The relay generates an estimate, $\hat j$, of $j$. In the
  second sub-slot, the source sends $\sqrt{\frac{\beta P_t}{2W}} v_k$
  and the relay sends $\sqrt{\frac{(1-\beta)P_t}{2W}} s_{\hat j}^2$,
  where $0 \leq \beta \leq 1$. Like before, $\alpha$ and $\beta$ are
  the power control functions depending on the link gain vector $Z$.
 
  As in part 2) above, when this code is used, the input-output
  relationship of the channel can be described by
  \begin{eqnarray*}
  Y^{n_1} &=&
  \sqrt{\frac{\alpha Z_{13} P_t}{2W}} \tilde X_1^{n_1} +
  \sqrt{\frac{(1-\alpha)Z_{13} P_t}{2W}} \tilde X_2^{n_1} + N^{n_1}
  \\
  Y_1^{n_1} &=& 
  \sqrt{\frac{\alpha Z_{12} P_t}{2W}} \tilde X_1^{n_1} +
  \sqrt{\frac{(1-\alpha) Z_{12} P_t}{2W}} \tilde X_2^{n_1} + N_1^{n_1}
  \end{eqnarray*}
  during the first sub-slot with $\tilde X_1^{n_1}$ corresponding to
  the codeword $u_i$, $\tilde X_2^{n_1}$ corresponding to the codeword
  $s^1_j$, and $X_1^{n_1} = \sqrt{\frac{\alpha P_t}{2W}} \tilde
  X_1^{n_1} + \sqrt{\frac{(1-\alpha) P_t}{2W}}\tilde X_2^{n_1}$ as the
  input to the CB channel. In the second sub-slot, we have
  \[
  Y^{n_2} = \sqrt{\frac{\beta Z_{13} P_t}{2W}} \tilde X_1^{n_2} +
  \sqrt{\frac{(1-\beta) Z_{23} P_t}{2W}} X_2^{n_2} + N^{n_2}
  \]
  with $\tilde X_1^{n_2}$ and $\tilde X_2^{n_2}$ corresponding to
  $v_k$ and $s^2_{\hat j}$, respectively.

  \paragraph{Decoding}
  We combine the decoding approaches suggested for the CB and MA
  channels in \cite{Boucheron00} and \cite{Han03}, respectively.  Fix
  any $\gamma >0$. Define the sets
  \begin{eqnarray*}
  T_1^{n}(\alpha,\beta,Z) &=&
    \left\{ \left(\tilde x_1^{n_1}, y^{n_1}\right) :
    \frac{1}{n} \log \frac{p_{Y^{n}|\tilde X_1^{n_1},\alpha,\beta,Z}
      (y^{n_1}|\tilde x_1^{n_1})}{p_{Y^{n_1}|\alpha,\beta, Z} (y^{n_1})} 
    > \frac{1}{n} \log M_n^1 + \gamma \right\} \\
  T_2^{n}(\alpha,\beta,Z) &=&
    \left\{ \left(\tilde x_1^{n_1},\tilde x_2^{n_1},y_1^{n_1}\right) :
    \frac{1}{n} \log \frac{p_{Y_1^{n_1}|\tilde X_1^{n_1}, \tilde X_2^{n_1},
        \alpha,\beta, Z}
      (y_1^{n_1}|\tilde x_1^{n_1},\tilde x_2^{n_1})}{
      p_{Y_1^{n_1}|\tilde X_1^{n_1},\alpha,\beta,Z}
      (y_1^{n_1}|\tilde x_1^{n_1})}
    > \frac{1}{n} \log M_n^2 + \gamma \right\} \\
  T_3^{n}(\alpha,\beta,Z) &=&
    \left\{ \left(\tilde x_1^{n_1},\tilde x_2^{n_1},y_1^{n_1}\right) :
    \frac{1}{n} \log \frac{p_{Y_1^{n_1}|\tilde X_1^{n_1},
        \tilde X_2^{n_1},\alpha,\beta, Z}
      (y_1^{n_1}|\tilde x_1^{n_1},\tilde x_2^{n_1})}{
      p_{Y_1^{n_1}|\alpha,\beta, Z} (y_1^{n_1})}
    > \frac{1}{n} \log M_n^1 M_n^2 + \gamma
  \right\} \\
  T_4^{n}(\alpha,\beta,Z) &=&
    \left\{ \left(\tilde x_1^{n_2},\tilde x_2^{n_2},y^{n_2}\right) :
    \frac{1}{n} \log \frac{p_{Y^{n_2}|\tilde X_1^{n_2},\tilde X_2^{n_2},
        \alpha,\beta, Z}(y^{n_2}|\tilde x_1^{n_2},\tilde x_2^{n_2})}%
    {p_{Y^{n_2}|\tilde X_2^{n_2},\alpha,\beta,Z}(y^{n_2}|\tilde x_2^{n_2})} 
    > \frac{1}{n} \log M_n^3 + \gamma \right\} \\
  T_5^{n}(\alpha,\beta,Z) &=&
    \left\{ \left(\tilde x_1^{n_2},\tilde x_2^{n_2},y^{n_2}\right) :
    \frac{1}{n} \log \frac{p_{Y^{n_2}|\tilde X_1^{n_2},\tilde X_2^{n_2},
        \alpha,\beta, Z}(y^{n_2}|\tilde x_1^{n_2},\tilde x_2^{n_2})}%
    {p_{Y^{n_2}|\tilde X_1^{n_2},\alpha,\beta,Z}(y^{n_2}|\tilde x_1^{n_2})}
    > \frac{1}{n} \log M_n^2 + \gamma \right\} \\
  T_6^{n}(\alpha,\beta,Z) &=&
    \left\{ \left(\tilde x_1^{n_2},\tilde x_2^{n_2},y^{n_2}\right) :
    \frac{1}{n} \log \frac{p_{Y^{n_2}|\tilde X_1^{n_2},\tilde X_2^{n_2},
        \alpha,\beta, Z}(y^{n_2}|\tilde x_1^{n_2},\tilde x_2^{n_2})}%
    {p_{Y^{n_2}|\alpha,\beta,Z} (y^{n_2})}
    > \frac{1}{n} \log M_n^2 M_n^3 + \gamma \right\}.
  \end{eqnarray*}
  In the first sub-slot, the relay sets $\hat j = j$ if and only if
  there is a unique pair $(i,j)$ such that $(u_i,s^1_j,y_1^{n_1}) \in
  T_2^{n}(\alpha,\beta,Z) \cap T_3^n(\alpha,\beta,Z)$. This allows the
  encoding step in the second sub-slot mentioned above. The
  destination outputs $i$ if and only if there is a unique $i$ such
  that $(u_i,y^{n_1}) \in T_1^{n}(\alpha,\beta,Z)$. In the second
  sub-slot, the destination outputs $(j,k)$ if there is a unique pair
  $(j,k)$ such that $(s^2_j,v_k,y^{n_2}) \in T_4^{n}(\alpha,\beta,Z)
  \cap T_5^n(\alpha,\beta,Z) \cap T_6^n(\alpha,\beta,Z)$.  Finally,
  the estimate of the message is then $\hat w = (i,j,k)$.

  \paragraph{Error analysis}
  Because of the symmetry of the code, we can assume $w=(1,1,1)$. Let
  $\varepsilon_{n}$ be the average error probability of decoding. For
  $i=1,2,\ldots,M_n^1$, write $E_{i}^1 = \left\{(u_i,y^{n_1}) \in
    T_1^{n}(\alpha,\beta,Z) \right\}$ and for $i=1,2,\ldots,M_n^1$ and
  $j=1,2,\ldots,M_n^2$, $E_{ij}^2 = \left\{(u_i,s^1_j,y_1^{n_1}) \in
    T_2^{n}(\alpha,\beta,Z) \cap T_3^n(\alpha,\beta,Z)\right\}$. For
  $j=1,2,\ldots,M_n^2$ and $k=1,2,\ldots,M_n^3$, $E_{jk}^3 =
  \left\{(v_k,s^2_j,y^{n_2}) \in T_4^{n}(\alpha,\beta,Z) \cap
    T_5^n(\alpha,\beta,Z) \cap T_6^n(\alpha,\beta,Z) \right\}$. Then
  \begin{equation}
  \varepsilon_n \leq \Pr \left( (E_1^1)^c \cup (E_{11}^2)^c \cup
    (E_{11}^3)^c \right) + \sum_{i=2}^{M_n^1} \Pr (E_i^1) +
  \sum_{(i,j) \neq (1,1)} \Pr (E_{ij}^2) + 
  \sum_{(j,k) \neq (1,1)} \Pr (E_{jk}^3).
  \label{e:hfd1err}
  \end{equation}
  Using the definitions of $T_1^n(\alpha,\beta,Z)$ to
  $T_6^n(\alpha,\beta,Z)$ and similar to part 2) (see
  \cite{Boucheron00,Han03} for the detailed arguments), one can show
  that the second, third, and fourth terms on the right hand side of
  (\ref{e:hfd1err}) can be bounded by $e^{-n\gamma}$, $2e^{-n\gamma}$,
  and $3e^{-n\gamma}$, respectively.

  As $n$ becomes large,
  \begin{eqnarray*}
    \frac{1}{n} \log \frac{p_{Y^{n}|\tilde X_1^{n_1},\alpha,\beta,Z}
      (y^{n_1}|\tilde x_1^{n_1})}{p_{Y^{n_1}|\alpha,\beta, Z} (y^{n_1})} 
    &\stackrel{\mathrm{a.s.}}{\longrightarrow}& 
    t_1 C\left(\frac{\alpha Z_{13} P_t}{(1-\alpha)Z_{13}P_t + N_0W}\right)
    \\
    \frac{1}{n} \log \frac{p_{Y_1^{n_1}|\tilde X_1^{n_1},
       \tilde X_2^{n_1},\alpha,\beta, Z}
      (y_1^{n_1}|\tilde x_1^{n_1},\tilde x_2^{n_1})}{
      p_{Y_1^{n_1}| \tilde X_1^{n_1},\alpha,\beta,Z}
      (y_1^{n_1}|\tilde x_1^{n_1})}
    &\stackrel{\mathrm{a.s.}}{\longrightarrow}& 
    t_1 C\left(\frac{(1-\alpha) Z_{12} P_t}{N_0W}\right)
    \\
    \frac{1}{n} \log \frac{p_{Y_1^{n_1}|\tilde X_1^{n_1},\tilde X_2^{n_1},
        \alpha,\beta, Z}(y_1^{n_1}|\tilde x_1^{n_1},\tilde x_2^{n_1})}%
        {p_{Y_1^{n_1}|\alpha,\beta, Z} (y_1^{n_1})}
    &\stackrel{\mathrm{a.s.}}{\longrightarrow}& 
    t_1 C\left(\frac{Z_{12} P_t}{N_0W}\right)
    \\
    \frac{1}{n} \log \frac{p_{Y^{n_2}|\tilde X_1^{n_2},\tilde X_2^{n_2},
        \alpha,\beta, Z}(y^{n_2}|\tilde x_1^{n_2},\tilde x_2^{n_2})}%
    {p_{Y^{n_2}|\tilde X_2^{n_2},\alpha,\beta,Z}(y^{n_2}|\tilde x_2^{n_2})} 
    &\stackrel{\mathrm{a.s.}}{\longrightarrow}& 
    t_2 C\left(\frac{\beta Z_{13} P_t}{N_0W}\right) 
    \\
    \frac{1}{n} \log \frac{p_{Y^{n_2}|\tilde X_1^{n_2},\tilde X_2^{n_2},
        \alpha,\beta, Z}(y^{n_2}|\tilde x_1^{n_2},\tilde x_2^{n_2})}%
    {p_{Y^{n_2}|\tilde X_1^{n_2},\alpha,\beta,Z}(y^{n_2}|\tilde x_1^{n_2})} 
    &\stackrel{\mathrm{a.s.}}{\longrightarrow}& 
    t_2 C\left(\frac{(1-\beta) Z_{23} P_t}{N_0W}\right) 
    \\
    \frac{1}{n} \log \frac{p_{Y^{n_2}|\tilde X_1^{n_2},\tilde X_2^{n_2},
        \alpha,\beta, Z}(y^{n_2}|\tilde x_1^{n_2},\tilde x_2^{n_2})}%
    {p_{Y^{n_2}|\alpha,\beta,Z} (y^{n_2})}
    &\stackrel{\mathrm{a.s.}}{\longrightarrow}& 
    t_2 C\left(\frac{\beta Z_{13} P_t + (1-\beta)Z_{23}P_t}{N_0W}\right) 
  \end{eqnarray*}
  when the inputs symbols are Gaussian distributed as described in the
  code generation step above.  For the cases of $t_1=0$ and $t_1=1$,
  the channel reduces to the case of MA and CB, respectively. Hence
  the corresponding subset of code construction should be employed.
  
  Now let $M_n^1=e^{nx_1}$, $M_n^2=e^{nx_2}$, and $M_n^3=e^{nx_3}$
  such that $x_1,x_2,x_3 \geq 0$ and $x_1+x_2+x_3=K$.  Choose $t_1$,
  $\alpha$, $\beta$, $x_1$, $x_2$, and $x_3$ as functions of $Z$ to
  minimize $P_t$. Since $\gamma>0$ is arbitrary, by
  Theorem~\ref{thm:b1}, the code construction above shows the
  existence of a sequence of $(n,M_n,\varepsilon_n,P_n)$-codes over
  the time slot with satisfying $\liminf_{n\rightarrow\infty}
  \frac{1}{n} \log M_{n} = K$, $\limsup_{n\rightarrow\infty}
  \varepsilon_{n} \leq P_1(K,S)$, and $\limsup_{n\rightarrow\infty}
  P_{n} = P_t$ a.s.
  
  Indeed, to see that Theorem~\ref{thm:b1} applies here, it suffices
  to show that the optimization solution described in
  Section~\ref{se:b1k} lies within the following region
  \begin{eqnarray*}
    t_1 C\left(\frac{\alpha Z_{13} P_t}{(1-\alpha)Z_{13}P_t + N_0W}\right) 
    &>& x_1
    \\
    t_1 C\left(\frac{(1-\alpha) Z_{12} P_t}{N_0W}\right)
    &>& x_2
    \\
    t_1 C\left(\frac{Z_{12} P_t}{N_0W}\right)
    &>& x_1+x_2
    \\
    t_2 C\left(\frac{\beta Z_{13} P_t}{N_0W}\right) 
    &>& x_3
    \\
    t_2 C\left(\frac{(1-\beta) Z_{23} P_t}{N_0W}\right) 
    &>& x_2
    \\
    t_2 C\left(\frac{\beta Z_{13} P_t + (1-\beta)Z_{23}P_t}{N_0W}\right) 
    &>& x_2+x_3.
  \end{eqnarray*}
  The last three inequality coincide with the MA region specified in
  part 2) of Lemma~\ref{thm:cbma} (see Appendix~\ref{app:cbma}). For
  $Z_{13} < Z_{12}$, it is easy to see the third inequality is
  redundant in place of the first two inequalities, which coincide
  with the CB region in part 1) of Lemma~\ref{thm:cbma}. For $Z_{13}
  \geq Z_{12}$, the optimal solution specified in
  Appendix~\ref{app:soln} can be achieved by the choice of $t_1=0$,
  hence making only the last three inequalities matter.
  
\item All the arguments are essentially the same as in part 3) with
  the modification that the source sends $\sqrt{\frac{\beta P_t}{2W}}
  v_k+ \sqrt{\frac{\delta (1-\beta)P_t}{2W}} s^2_j$ and the relay
  sends $\sqrt{\frac{(1-\delta)(1-\beta)P_t}{2W}} s^2_{\hat j}$, where
  $0 \leq \delta \leq 1$ is an additional power control component, in
  the second sub-slot.
\end{enumerate}

\subsection{Proof of Theorem~\ref{thm:dlc}}
\label{app:dlc}
We sketch the proof of the theorem, which employs results from
\cite{Caire99} directly.  Below we use the index $i$ to denote one of
the four cases of lower bound $(i=\mathrm{lb})$, decode forward
$(i=\mathrm{DF})$, HDP1 $(i=1)$, and HDP2 $(i=2)$.

For $i\in \{\mathrm{DF},1,2\}$, replacing $P_t$ by $S(Z)[e^K-1]N_0W$
in the proofs parts 2) -- 4) of Theorem~\ref{thm:ec} given in
Appendix~\ref{app:ec}, we can show the existence of a sequence of
$(n,e^{nK},\varepsilon_n,P_n)$-codes with $P_n \leq S(Z)[e^K-1]N_0W+
\gamma$ a.s. and $\varepsilon_n \leq P_i(K,S(Z))+\gamma$, for any
$\gamma>0$, whenever $n$ is sufficiently large. Note that we write the
RNSNR $S(Z)$ to highlight the use of a general power control scheme
which varies the total transmit energy (rather than setting it to a
fixed value as in the original proofs) in each time slot according to
the link gains.

Consider the optimal power control function $\hat S_i(Z)$ that solves
the following optimization problem:
\begin{equation}
\begin{array}{ll}
\min_{S(Z)} & P_i(K,S(Z)) \\ 
\mbox{subject~to} & E[S(Z)] 
\leq \bar S \stackrel{\triangle}{=} \frac{P_t}{N_0W (e^K-1)}.
\end{array}
\label{e:optlt}
\end{equation}
Write $B_{\mathrm{lb}}$, $B_{\mathrm{DF}}$, $B_1(K)$, and $B_2(K)$ as
$B_i(Z)$ for $i= \mathrm{lb},\mathrm{DF},1$, and $2$, respectively to
highlight their dependence on $Z$. Define $\mathcal{S}_i(s) =
\int_{\{Z:B_i(Z) \leq s\}} B_i(Z) dF_Z$, where $F_Z$ is the
distribution function of the link gain vector $Z$. Let $s^*_i = \sup
\{s: \mathcal{S}_i(s) < \bar S\}$. Then Proposition~4 of
\cite{Caire99} can be applied to solve the optimization problem in
(\ref{e:optlt}), provided that $B_i(Z)$ is continuous for all $Z\neq
0$ and is non-increasing in each of the elements of $Z$ (c.f. Lemma~2
of \cite{Caire99}). This latter condition is established in
Corollaries~\ref{thm:b1asymp} and \ref{thm:b2asymp} for $i=1$ and $2$,
respectively. For $i=\mathrm{lb}$ and $\mathrm{DF}$, the condition can
be checked by straightforward calculus. The resulting solution is
\[
\hat S_i(Z) = \left\{
  \begin{array}{ll}
    B_i(Z) & \mbox{~if~} B_i(Z) < s^*_i \\
    0 & \mbox{~otherwise}
  \end{array} \right.
\]
and $P_i(K,\hat S_i(Z)) = \Pr (B_i(Z) \geq s^*_i)$.  Now, if
$E[B_i(Z)] = \lim_{s\rightarrow\infty} \mathcal{S}_i(s)$ is finite,
setting $\bar S = E[B_i(Z)]$ will make $s^*_i = \infty$ and hence
$P_i(K,\hat S_i(Z)) = 0$ as well as $E[\hat S_i(Z)] = E[B_i(Z)]$.

For the cases of $i=\mathrm{DF},1$, and $2$, this implies that the
rate $K$ is long-term $E[B_{\mathrm{DF}}(Z)](e^K-1)N_0W$-achievable,
long-term $E[B_1(Z)](e^K-1)N_0W$-achievable with HDP1, and long-term
$E[B_2(Z)](e^K-1)N_0W$-achievable with HDP2, respectively.  For
$i=\mathrm{lb}$, a slight modification to the proof of part 1) of
Theorem~\ref{thm:ec} shows that if the rate $K$ is long-term $\bar
S(e^K-1)N_0W$-achievable, then the error probabilities of the sequence
of codes (and the corresponding power control schemes $S_n(Z)$) must
satisfy $E[S_n(Z)] \leq \bar S + \gamma$ and $\delta P_{\mathrm{lb}}
\left(K, \frac{e^K-1}{e^{(1-\delta) K} - 1} S_n(Z) \right) \leq
\varepsilon_n < \gamma $ for all $0<\delta<1$ and any $\gamma >0$,
whenever $n$ is sufficiently large.  Since $P_{\mathrm{lb}}$ is
continuous in the second argument, this requires that
$P_{\mathrm{lb}}(K,S_n(Z))=0$ for all sufficiently large $n$. The
solution of (\ref{e:optlt}) then implies that $\bar S \geq
E[B_{\mathrm{lb}}(Z)]$.

Since $B_{\mathrm{lb}}(Z) \leq B_{\mathrm{DF}}(Z) \leq B_2(Z) \leq
B_1(Z)$ for all $Z$, it suffices to establish the finiteness of
$E[B_1(Z)]$. From the proof of part 3) of Theorem~\ref{thm:outprob},
\[
B_1(Z) \leq \hat B_1(Z) \stackrel{\triangle}{=} \left\{
  \begin{array}{ll}
    \displaystyle
    \sqrt{\frac{2}{Z_{13}}} \sqrt{\frac{1}{Z_{12}} +\frac{1}{Z_{23}}}
      &
    \mbox{~if~} Z_{13} < M_H(Z_{12},Z_{23})\\
    \displaystyle
    \frac{1}{Z_{13}} &
    \mbox{~if~} Z_{13} \geq M_H(Z_{12},Z_{23}).
  \end{array} \right.
\]  
Thus it in turn suffices to establish the finiteness of $E[\hat
B_1(Z)]$. Indeed
\begin{eqnarray}
  E[\hat B_1(Z)] &=& 
  \int_{\frac{1}{x} < \frac{1}{y}+\frac{1}{z}} \frac{1}{x} e^{-(x+y+z)} dxdydz
    + \int_{\frac{1}{x} \geq \frac{1}{y}+\frac{1}{z}} 
      \sqrt{\frac{2}{x}}\cdot\sqrt{\frac{1}{y}+\frac{1}{z}} e^{-(x+y+z)} dxdydz
   \nonumber\\
 &=&
 \underbrace{\int_{0}^{\infty} \frac{1}{x} 
   \left[ 1 - 2xe^{-2x}K_1(2x)\right] e^{-x} dx}_{a}
 \nonumber \\
 & & ~~~
 + \underbrace{\int_{0}^{\infty} \sqrt{\frac{2}{x}} e^{-x} \int_{x}^{\infty}
   2 \sqrt{u} e^{-2u} \left[K_0(2u)+ K_1(2u)\right] dudx}_{b}
 \label{e:ehb1}
\end{eqnarray}
where the second equality is obtained by using the integral
representations of $K_0(x)$ and $K_1(x)$ in \cite[pp. 969]{Grad94}.
Again using the property of the modified Bessel functions, it is easy
to check that the integrand in the integral $a$ on the right hand side
of (\ref{e:ehb1}) is bounded above over the range of $0 \leq x \leq 1$
and is bounded above by $e^{-x}$ for $x > 1$. Thus $a$ is finite. On
the other hand, we have
\[
b \leq 
 \left(\int_{0}^{\infty} \sqrt{\frac{2}{x}} e^{-x} dx\right) \cdot
 \left( \int_{0}^{\infty}
   2 \sqrt{u} e^{-2u} \left[K_0(2u)+ K_1(2u)\right] du \right)
\]
where the first integral on the right hand side is
$\sqrt{2}\Gamma\left(\frac{1}{2}\right)$ (see \cite[pp. 942]{Grad94})
and the second integral is finite (see \cite[pp. 733]{Grad94}). Thus
$b$ is also finite.

\clearpage

\begin{figure}
  \centering \includegraphics[width=\textwidth]{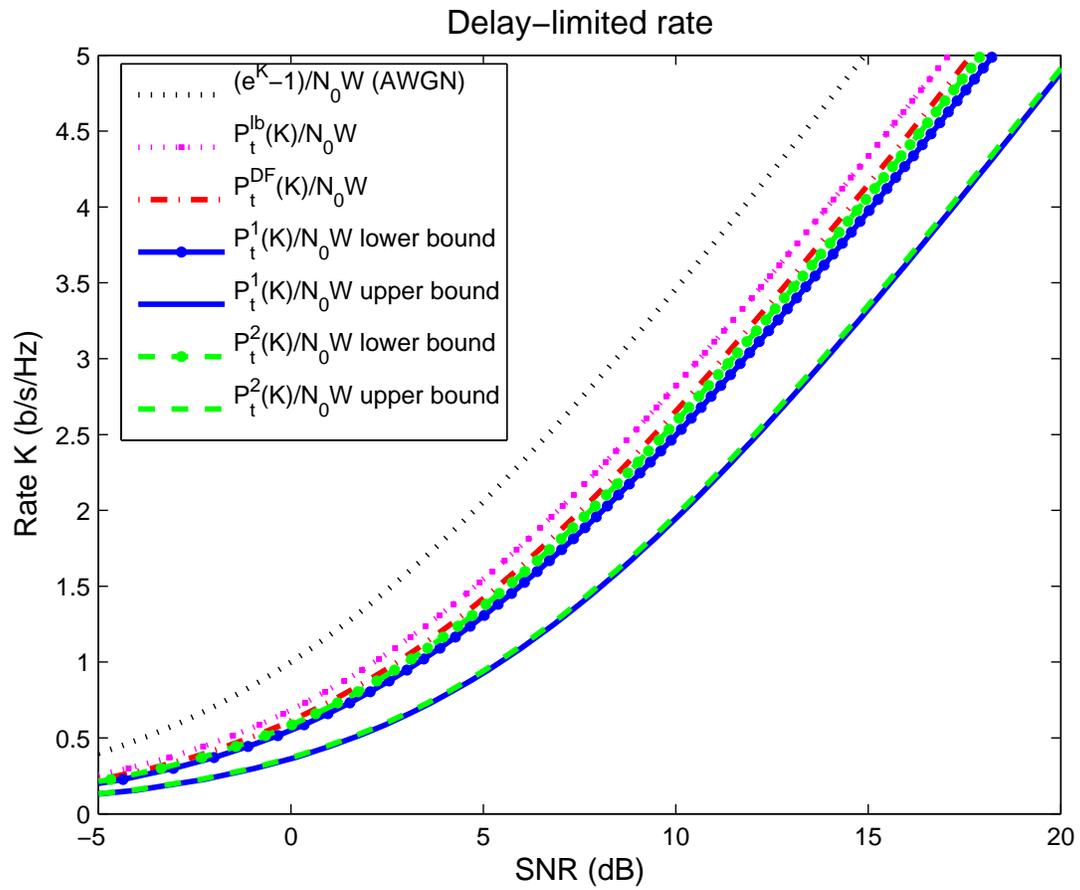}
  \caption{Plot of delayed-limited rates for various transmission schemes.}
  \label{f:dlrate}
\end{figure}

\begin{table}
  \centering
  \caption{Decibel losses in performance with respective to an AWGN channel.}
  \label{tb:loss}
  {\large
  \begin{tabular}{|c|c|c|c|c|c|}
  \hline
    $E[B_{\mathrm{lb}}]_{\mathrm{dB}}$ & 
    $E[B_{\mathrm{DF}}]_{\mathrm{dB}}$ & 
    $E[B_{1}(0)]_{\mathrm{dB}}$ & 
    $E[B_{1}(\infty)]_{\mathrm{dB}}$ & 
    $E[B_{2}(0)]_{\mathrm{dB}}$ & 
    $E[B_{2}(\infty)]_{\mathrm{dB}}$ 
    \\ \hline
    $2.17$ & $2.76$ & $3.33$ & $5.45$ & $3.02$ & $5.36$ \\
    \hline
  \end{tabular}}
\end{table}

\end{document}